\documentclass[1p,preprint,10pt]{elsarticle}

\usepackage{amsmath}
\usepackage{amsfonts}
\usepackage{array}
\usepackage{xcolor}
\usepackage{xspace}
\usepackage{hyperref}
\usepackage{lineno}
\usepackage{listings}
\usepackage{tabularx}
\usepackage{tabulary}
\usepackage{makecell}
\usepackage{placeins}
\usepackage{graphicx}
\usepackage{amssymb}
\usepackage{todonotes}
\usepackage{bbold}
\definecolor{linkcolor}{HTML}{399B03}
\definecolor{urlcolor}{HTML}{399B03}
\hypersetup{pdfstartview=FitH, linkcolor=linkcolor,urlcolor=urlcolor,colorlinks=true}
\hypersetup{
    unicode=false,          
    pdftoolbar=true,        
    pdfmenubar=true,        
    pdffitwindow=false,     
    pdfstartview={FitH},    
    pdfauthor={},         
    colorlinks=true,        
    linkcolor=blue,         
    citecolor=red,          
}

\newcommand{\TRIQS}{\textnormal{\texttt{TRIQS}}\xspace}
\newcommand{\Nevanlinna}{\textnormal{\texttt{Nevanlinna}}\xspace}

\definecolor{darkblue}{rgb}{0,0,.6}
\definecolor{darkred}{rgb}{.6,0,0}
\definecolor{darkgreen}{rgb}{0,.6,0}
\definecolor{red}{rgb}{.98,0,0}

\def\ssmall{\fontsize{8pt}{2pt}\selectfont}

\lstnewenvironment{pylisting}[1][]
{\lstset{language=Python,numbers=none,#1}}{}

\journal{Computer Physics Communications}

\begin{document}

\begin{frontmatter}

\title{\TRIQS/\Nevanlinna: Implementation of the Nevanlinna Analytic Continuation method for noise-free data.}

\author[Michigan]{Iskakov Sergei\corref{author}}
\ead{siskakov@umich.edu}
\author[NewYork]{Alexander Hampel}
\ead{ahampel@flatironinstitute.org}
\author[NewYork]{Nils Wentzell}
\ead{nwentzell@flatironinstitute.org}
\author[Michigan]{Emanuel Gull}
\ead{egull@umich.edu}

\cortext[author] {Corresponding author. Current address: Department of Physics,
University of Michigan, Ann Arbor, Michigan 48109, USA}

\address[Michigan]{Department of Physics, University of Michigan, Ann Arbor, Michigan 48109, USA}
\address[NewYork]{Center for Computational Quantum Physics, Flatiron Institute, 162 5th Avenue, New York, NY 10010, USA}

\begin{abstract}
We present the \TRIQS/\Nevanlinna analytic continuation package, an efficient
implementation of the methods proposed by J. Fei \textit{et al} in [Phys. Rev. Lett. \textbf{126}, 056402 (2021)] and [Phys. Rev. B \textbf{104}, 165111 (2021)].
\TRIQS/\Nevanlinna strives to provide a high quality open source
(distributed under the GNU General Public License version 3)
alternative to the more widely adopted Maximum Entropy based analytic continuation
programs. With the additional Hardy functions optimization procedure, it allows for an
accurate resolution of wide band and sharp features in the spectral function.
Those problems can be formulated in terms of imaginary time or Matsubara frequency response functions. The application is based on the \TRIQS C++/Python
framework, which allows for easy interoperability with other \TRIQS-based applications, electronic band  structure codes and visualization tools.
Similar to other \TRIQS packages, it comes with a convenient Python interface.
\end{abstract}

\end{frontmatter}

\noindent {\bf PROGRAM SUMMARY}

\begin{small}
\noindent
{\em Program Title:} \TRIQS/\Nevanlinna \\
{\em Developer's repository link:} \url{https://github.com/TRIQS/Nevanlinna} \\
{\em Programming language:} \verb*#C++#/\verb*#Python#\\
{\em Licensing provisions:} GNU General Public License (GPLv3)
{\em External routines/libraries:}
\verb#TRIQS 3.2#\cite{TRIQS}, \verb#Boost >= 1.76.0#, \verb#Eigen >= 3.4.0#, \verb#cmake >= 3.20 #.\\
{\em Nature of problem:}
Finite-temperature field theories are widely used
to study quantum many-body effects and electronic structure of correlated
materials. Obtaining physically relevant spectral functions from results in the imaginary time/Matsubara frequency domains
requires solution of an ill-posed analytic continuation problem as a post-processing step.\\
{\em Solution method:}
We present an efficient C++/Python open-source implementation of the Nevanlinna/Caratheodory analytic continuation.
\end{small}

\section{\label{sec:introduction}Introduction}
The single-particle spectral function of interacting quantum many-body systems is an object of prime importance in many-body theory, since it reveals excitation information of a quantum system. This information can often be directly compared to experiment. Up to a prefactor, the spectral function corresponds to the imaginary part of the retarded real-frequency Green's function \cite{Mahan00}. Finite-temperature field-theoretical methods, such as path-integral \cite{Ceperley95} and lattice quantum Monte Carlo \cite{Blankenbecler81,Scalapino81}, self-consistent finite-temperature perturbation theory \cite{Hedin65,Holleboom90,Dahlen06,Stan09,Phillips14,Rusakov16,Yeh22}, non-perturbative embedding theories \cite{Metzner89,Georges92,Georges96,Kotliar06,Zgid17,Rusakov19}, diagrammatic Monte carlo methods \cite{Prokofev98,Prokofev07} or continuous-time Monte Carlo impurity solvers \cite{Rubtsov05,Werner06,Gull08,Gull11} provide direct access to the so-called imaginary-time or Matsubara frequency Green's functions, which are closely related to the retarded real-frequency Green's function. The problem of numerical analytic continuation then consists of obtaining retarded real-frequency Green's functions from the Matsubara and imaginary time quantities obtained in the aforementioned computational methods. A naive solution of the analytic continuation equations is ill conditioned, in the sense that a small change of the input Matsubara quantities results in a large change of the resulting spectral function \cite{Jarrell96}.

In the presence of noise, numerous numerical approaches, including the Maximum Entropy method \cite{Jarrell96,Bergeron16,Levy17,Huang22,Kaufmann23} and variants \cite{Burnier13}, the stochastic analytic continuation \cite{Sandvik98,Mishchenko00,Fuchs10,Sandvik16,Krivenko19,Shao23} and variants \cite{Goulko17}, as well as sparse modeling \cite{Otsuki17,Yoshimi19,Otsuki20}, the Prony method \cite{Ying22}, machine learning \cite{Yoon18,Fournier20}, and projection, pole estimation, and semidefinite relaxation approaches \cite{Huang23}, and Pad\'{e} approximants \cite{Vidberg77,Beach00} have been successful. Extensions and generalizations have also appeared to the problem of matrix-valued spectral functions \cite{Kraberger17}, non-positive spectral functions \cite{Rothkopf17} and superconductivity \cite{Gull14,Reymbaut15,Dong22,Yue23}.

In the absence of noise, an interpolation of the input data with an appropriate function, rather than a fit, is desired.
Apart from coinciding with input values, the interpolating function should also respect the analytic properties of Green's
functions, in particular causality. The space of scalar causal response functions is spanned by so-called Nevanlinna
functions~\cite{Fei21,Nevanlinna25}, while their matrix-valued generalizations are known as Carath\'{e}odory
functions~\cite{Fei21A,Caratheodory1907} (in other contexts, such functions are also known as Herglotz, Pick, or Riesz  functions).
This function space is very restrictive, in the sense that physical response functions only exist for data that fulfills
the so-called Pick criterion~\cite{Pick17}. A closed-form continued fraction interpolation algorithm that parametrizes all
possible causal response functions in terms of a free Nevanlinna function has been provided by Schur~\cite{Schur18}.

Nevanlinna analytic continuation by now has found applications in domains ranging from real-materials simulations~\cite{Yeh22,Yeh22B}
to high energy physics \cite{Bergamaschi23}. Generalizations to situations with noise~\cite{Huang23}, to bosonic response
functions~\cite{Nogaki23B,Bergamaschi23}, and open-source implementations of the Nevanlinna method Julia have appeared~\cite{Nogaki23}.
Nevertheless, there is a need for a generic, well-documented and efficient implementation of the method for scalar and matrix-valued
response functions, especially in the situation for continuous spectral functions where a function optimization is required to obtain
smooth spectra. This paper provides such an implementation based on the TRIQS software library for interacting quantum systems,
designed as a fully integrated application in the TRIQS ecosystem, but also usable standalone.

\section{\label{sec:theory}Theory}
We start the method description by introducing the fermionic imaginary time Green's function
\begin{align}
G_{\sigma,ij}(\tau) = -\langle T_{\tau} c_{\sigma,i}(\tau)c_{\sigma,j}^{\dagger}(0)\rangle,
\end{align}
with orbital index $i(j)$, spin index $\sigma$ and imaginary time $\tau$. The corresponding Matsubara Green's function can be obtained via
Fourier transform
\begin{align}
G_{\sigma,ij}(\omega_n) = \int_{0}^{\beta}G_{\sigma,ij}(\tau) e^{i\omega_n\tau},
    \label{eqn:greensfunction}
\end{align}
with fermionic Matsubara frequency $\omega_n = (2n+1)\frac{\pi}{\beta}$ for inverse temperature $\beta$.

The connection between Matsubara ($G_{\sigma,ij}(\omega_n)$) and retarded ($G_{\sigma,ij}^{R}(\omega)$) Green's functions is known as an analytical continuation.
If the analytical expression for the Matsubara Green's function is known everywhere in the complex plane, e.g.\ through the Lehmann representation,
the retarded Green's function can be obtained by the substitution $i\omega_n \rightarrow \omega + i\delta$ (where $\delta$ is a positive infinitesimal number).
In practice, the function is only known on the Matsubara axis, and obtaining the real frequency information requires solution
of an inhomogeneous Fredholm integral equation of the first kind with reference to the retarded Green's function
and known Matsubara Green's function on the left hand side:
\begin{align}
G_{\sigma,ij}(\omega_n) = -\frac{1}{\pi}\int_{-\infty}^{\infty}\frac{G^{R}_{\sigma,ij}(\omega)}{i\omega_n - \omega}d\omega.
\end{align}
The numerical solution if this equation is ill-posed~\cite{Jarrell96} and unstable in the presence of numerical noise.

In this work we present a C++ software package that solves a Nevanlinna-Pick problem~\cite{Schur18} using Nevanlinna
and Carath\'{e}odory functions for the scalar- and matrix-valued analytical continuation problem as introduced
in~\cite{Fei21,Fei21A}.

\subsection{Schur algorithm}\label{subsec:theory-schur}

The main mathematical background for Nevanlinna and Carath\'{e}odory analytical continuation methods is based on
the Schur algorithm. The Schur algorithm solves the Nevanlinna-Pick interpolation problem for contractive functions that are
analytic and bounded by 1 on the unit disk~\cite{Schur18}. For a given Schur function $S(z)$ we have a set of $M$ pairs
$(z_i, F_i)$ of node points $z_i$ and Schur parameters $F_i = S_i(z_i)$, that forms the following recurrence relation~\cite{43cc8a3d-e842-3b2d-b3bd-f44652881afb,doi:10.1137/0602013,CHEN1994253,10.1063/1.4826042}
\begin{align}
    S_{i+1}(z) = \frac{|z_i|(z_i - z)}{z_i(1 - z^{*}_i z)}
    \left[\mathbb{1} - F_i F^{\dagger}_i\right]^{-\frac{1}{2}}
    \left[S_{i}(z) - F_{i}\right]
    \left[\mathbb{1} - F^{\dagger}_{i} S_{i}(z)\right]^{-1}
    \left[\mathbb{1} - F^{\dagger}_i F_i\right]^{\frac{1}{2}},
    \label{eqn:schur}
\end{align}
here $X^{\frac{1}{2}}$, in case of a matrix-valued functions, standing for the Hermitian square root of X, and algebraic
square root for scalar-valued functions.

An iterative algorithm first constructs all Schur functions $S_i(z_j)$ at all node points $z_j$, then solves inverse problem
of Eq.~\ref{eqn:schur} for $z\in\mathcal{D}\setminus\{z_i\}$: $S_{M}(z) \rightarrow S_{M-1}(z) \rightarrow \ldots \rightarrow S_{0}(z) = S(z)$,
where $S(z)$ is the desired Schur interpolant and $S_{M}(z)$ is an arbitrary Schur function~\cite{doi:10.1137/0602013}.

\subsection{Nevanlinna continuation}\label{subsec:theory-nevanlinna}
Nevanlinna functions~\cite{nevanlinna1929beschrankte,akhiezer2020classical} belong to a class of complex functions which are analytic in the open upper half-plane $\mathcal{C}^{+}$
 and have non-negative imaginary part. Using the invertible M\"obius transform Nevanlinna functions $N(z)$ can be mapped onto a class of contractive functions
$\theta(z) = \frac{N(z) - i}{N(z) + i}$ that can be solved by Nevanlinna-Pick interpolation algorithm~\cite{Schur18,10.1155/S1687120003212028}.

In case of the scalar-valued functions, inverse recurrent problem of Eq.~\ref{eqn:schur} can be simplified and written as:
\begin{align}
    \theta_{0}(z) = \frac{a(z) \theta_{M}(z) + b(z)}{c(z) \theta_{M}(z) + d(z)},
\end{align}
where $\theta_{M}(z)$ is an arbitrary contractive function, $a(z), b(z), c(z)$ and $d(z)$ defined as~\cite{10.1155/S1687120003212028}
\begin{align}
    \left(\begin{matrix}a(z) & b(z)\\ c(z)& d(z)\end{matrix}\right) = \prod_{i=0}^{M-1}
    \left(\begin{matrix} \frac{z-Y_j}{z-Y^{*}_j} & \theta_{j}(z_j) \\ \theta^{*}_{j}(z_j)\frac{z - Y_j}{z-Y^{*}_j} & 1  \end{matrix}\right),
\end{align}
here $z\in\mathcal{C}^{+}$ and $Y_j$ are values of node points in $\mathcal{C}^+$ where the target Nevanlinna function has known values.

As it is shown in Ref.~\cite{Fei21}, a negative of the diagonal element of the fermionic Green's function, Eq.~\ref{eqn:greensfunction},
is a Nevanlinna function. With that we can construct Nevanlinna-Pick interpolation problem for a Nevanlinna function $N(z) = -G(z)$, using
positive Matsubara frequencies as node points.
After the Nevanlinna-Pick interpolation is performed, the Green's function on the real frequency can be obtained through the inverse M\"obius
transform:
\begin{align}
    G(z = \omega+i\eta) = - i \frac{1 + \theta(z)}{1 - \theta(z)}.
\end{align}

\subsection{Carath\'{e}odory continuation}\label{subsec:theory-caratheodory}
A matrix-valued function $F(w)$ defined on open subset $\mathcal{B}$ belongs to the class of Carath\'{e}odory
functions $\mathbb{C}$ if, for any $w \in \mathcal{B}$, the Hermitian matrix $\frac{1}{2}(F(z) + F^{\dagger}(z))$ is
positive semi-definite. Caratheodory functions $C(z)$ defined on a unit disk
$\mathcal{D} = {z : |z| < 1}$ can be mapped to corresponding Schur function using a Cayley transform:
\begin{align}
    S(z) = \left[\mathbb{1} - C(z)\right]\left[\mathbb{1} + C(z)\right]^{-1},
\end{align}
and the Nevanlinna-Pick interpolation can be constructed~\cite{doi:10.1137/0602013}.

It is easy to show that the product of a matrix-valued Green's function, Eq.~\ref{eqn:greensfunction}, and imaginary unit $i$
belongs to the class of Carath\'{e}odory functions~\cite{Fei21A}.
To solve Nevanlinna-Pick interpolation problem for matrix-valued Green's function, we first need to transform Green's function $G(w)$
into a Carath\'{e}odory function $C(w)$ and map it onto $\bar{C}(w)$ defined on a unit disk $\mathcal{D} = {z : |z| < 1}$ using M\"obius transform.
Then, using Cayley transform, we map this Carath\'{e}odory function to a corresponding Schur function $S(z)$ and perform
Schur algorithm (Eq.~\ref{eqn:schur}).
After the interpolation on the entire unit disk is obtained we subsequently apply inverse Cayley and M\"obius transform
to obtain Carath\'{e}odory function in the upper half of the complex plane
\begin{align}
    C(w) = \left[\mathbb{1} + \bar{C}(\frac{w-i}{w+i})\right]^{-1}\left[\mathbb{1} - \bar{C}(\frac{w-i}{w+i})\right].
\end{align}
The real frequency matrix-valued Green's function is then obtained as:
\begin{align}
    G(w = \omega+i\eta) = -i C(w).
\end{align}

\subsection{Existence criteria}\label{subsec:theory-pick}
Even though, mathematically both scalar- and matrix-valued Green's functions belong to Nevanlinna and Carath\'{e}odory
function up to a corresponding  prefactors ($-1$ and $i$), in actual calculations, due to round off errors, noise, and systematic approximation error
the restrictions imposed on Green's functions can be violated. However, there are straightforwardly verifiable criteria
for the existence of Nevanlinna and Carath\'{e}odory interpolants directly based on input data, which is a generalization of
the Pick criterion. An interpolant exists if and only if the corresponding Pick's matrix is positive semi-definite~\cite{Pick17,nevanlinna1929beschrankte}.
The Pick's matrix for Nevanlinna function is
\begin{align}
    P^{N}_{ij} = \left[\frac{1-\theta(Y_i)\theta^{*}(Y_j)}{1-\frac{Y_i- i}{Y_i +i}(\frac{Y_j- i}{Y_j +i})^*}\right]_{ij},
\end{align}
where $\theta(Y_i)$ is the corresponding value of contractive function at the Matsubara frequency $Y_i$.
Similarly, the Pick matrix for Caratheodory function is~\cite{10.1063/1.4826042,CHEN1994253}
\begin{align}
    P^{C} = \left[\frac{1-\bar{C}^{*}(z_j)\bar{C}(z_i)}{1-z_j^*z_i}\right]_{mn\times mn},
\end{align}

More detailed discussion of the underlying mathematics and the numerical procedure is presented in Refs.~\cite{Fei21,Fei21A}.

\section{Usage}\label{sec:usage}

\subsection{Installation}\label{subsec:installation}
The current version has the following dependencies: \TRIQS library version 3.2, Boost-Multiprecision library version 1.7.4 or higher,
Eigen3 library version 3.4.0 or higher and optionally MPFR library for fast multiprecision evaluation. The installation instructions of TRIQS are  provided on the website \url{https://triqs.github.io}.

Installing \Nevanlinna is similar to that of other \TRIQS-based applications.
Assuming that \TRIQS 3.2 has been installed at \verb|/path/to/TRIQS/install/dir|
\Nevanlinna is simply installed by issuing the following commands at the shell prompt:

\begin{verbatim}
$ git clone https://github.com/TRIQS/Nevanlinna
$ cd Nevanlinna
$ mkdir build && cd build
$ cmake -DTRIQS_PATH=/path/to/TRIQS/install/dir \
        -DCMAKE_INSTALL_PREFIX=/path/to/Nevanlinna/install/dir ../
$ make
$ make test
$ make install
\end{verbatim}
This will compile the source code, run automatic tests, and install \Nevanlinna in the \verb|/path/to/Nevanlinna/install/dir| directory. Further installation
instructions, such as installation without \TRIQS support, are given in the online documentation \cite{TriqsNevDocu}.

\subsection{Basic usage}\label{subsec:basic}
Running \Nevanlinna to analytically continue input data requires writing a simple
Python script. This usage method is standard for \TRIQS applications.
Prior to executing the script the user should load the \TRIQS as well as the
\TRIQS/\Nevanlinna installation into their current environment:
\begin{verbatim}
source /path/to/TRIQS/install/dir/share/triqs/triqsvars.sh
source /path/to/Nevanlinna/install/dir/share/nevanlinna/nevanlinnavars.sh
\end{verbatim}

Details of the script will vary depending on the physical observable to be continued,
and its representation. Nonetheless, a typical script will have the following basic
parts.

\begin{itemize}
    \item Import \TRIQS and \Nevanlinna Python modules.
    \begin{pylisting}
        # TRIQS Green's functions used to store input and output data.
        from triqs.gf import *
        # Use TRIQS HDF5 library for filesystem storage
        from h5 import HDFArchive

        # Import Nevanlinna solver
        from triqs_Nevanlinna import Solver
    \end{pylisting}

    \item Load the observable to be continued from an HDF5 archive.
    \begin{pylisting}
        # Open an HDF5 file and read Matsubara Green's function.
        ar = HDFArchive('input.h5', 'r')
        G_im = ar['input/gf']
        # or create a Gf input Green's function
        G_im = ...
    \end{pylisting}
    Only the values stored in the \lstinline[language=Python]{G_im.data} \verb#numpy# array, the TRIQS Green's function container data accessor, will be used by \Nevanlinna.
    \item Initialize the continuation solver
    \begin{pylisting}
        # Open an HDF5 file and read Matsubara Green's function.
        solver = Solver(kernel=<Kernel Type>, precision=<precision>)
    \end{pylisting}
    There are two possible options for kernel type: "NEVANLINNA" and "CARATHEODORY". By default, the Nevanlinna kernel will be used. One can also provide the desired precision for multi-precision operations if the \Nevanlinna code has
    been built with MPFR library support. By default, \lstinline[language=Python]{precision = 100} decimal digits is used.

    \item Solve the factorization problem for the desired kernel:
    \begin{pylisting}
        solver.solve(G_im)
    \end{pylisting}
    In current version, \lstinline[language=Python]{G_im} should be matrix-valued \TRIQS Green's function.
    If the Nevanlinna kernel is chosen the analytic continuation will only be performed for the 
    diagonal matrix elements and a warning message will be shown. To continue a full matrix valued Green's function including off-diagonal elements the Carath\'{e}odory kernel should be chosen.

    \item Setup the real-frequency grid and evaluate continuation.
    \begin{pylisting}
        # Setup equidistant grid and choose broadening parameter
        m = MeshReFreq(-5.0, 5.0, 101)
        eta = 0.1
        # Evaluate continuation for chosen grid and broadening eta
        G_re= solver.evaluate(m, eta)
    \end{pylisting}

    This function call is normally the most expensive part of the script. The result of the continuation will be stored in the \lstinline[language=Python]{G_re} container as a \TRIQS Green's function object.

\end{itemize}
\subsubsection{Parallelization}
\begin{figure}
    \includegraphics[width=.47\textwidth]{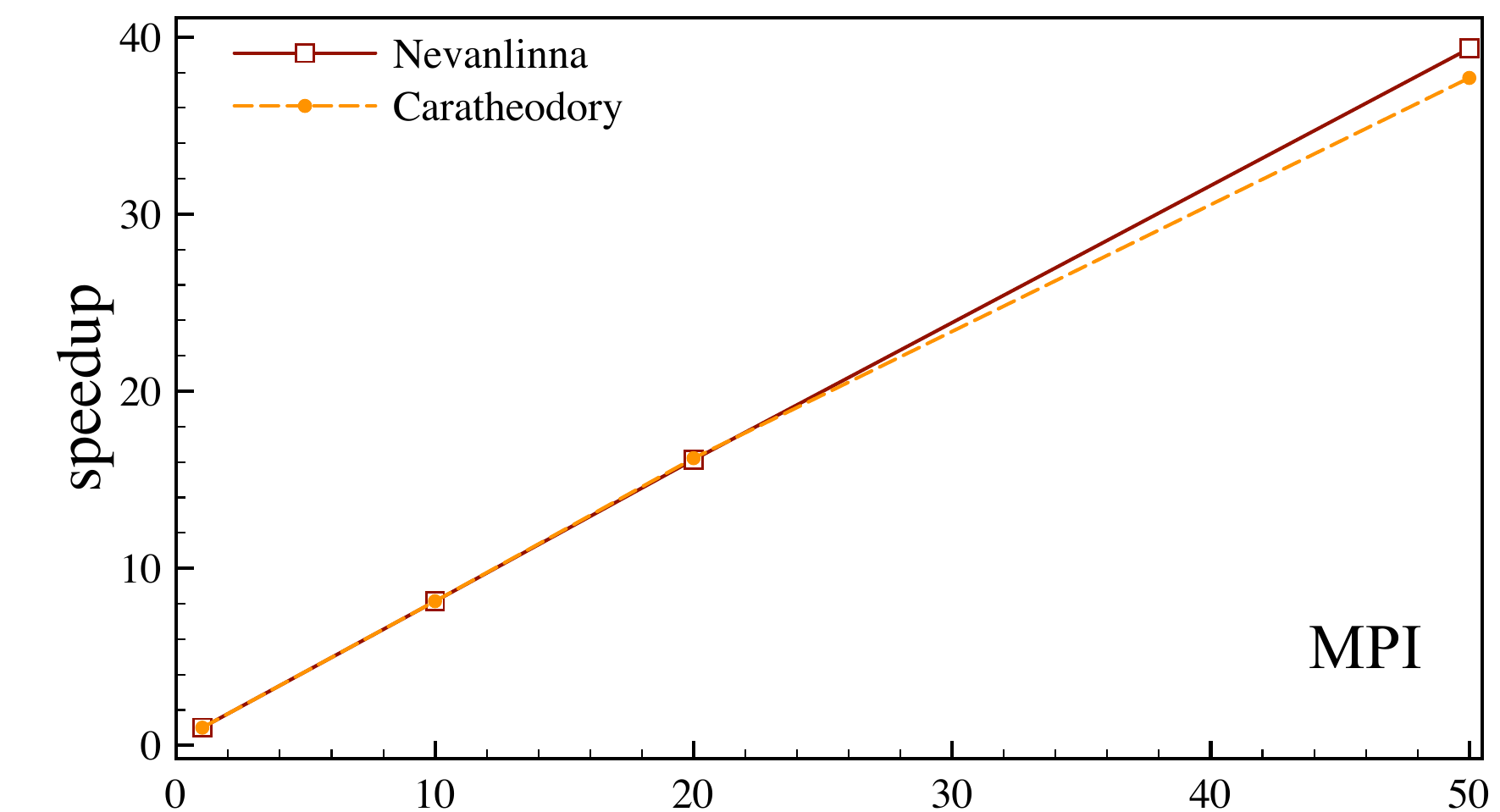}
    \includegraphics[width=.47\textwidth]{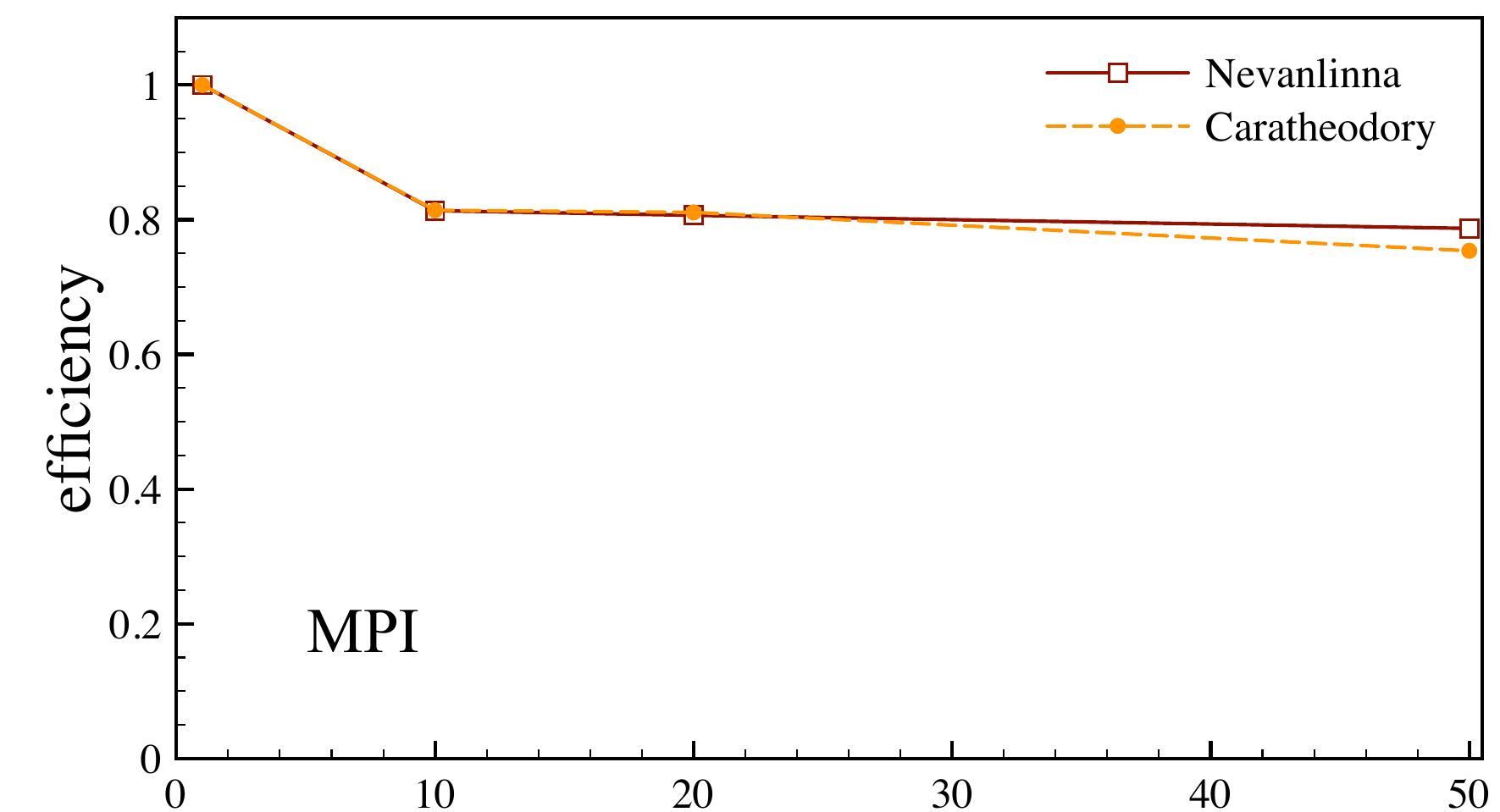}

    \includegraphics[width=.47\textwidth]{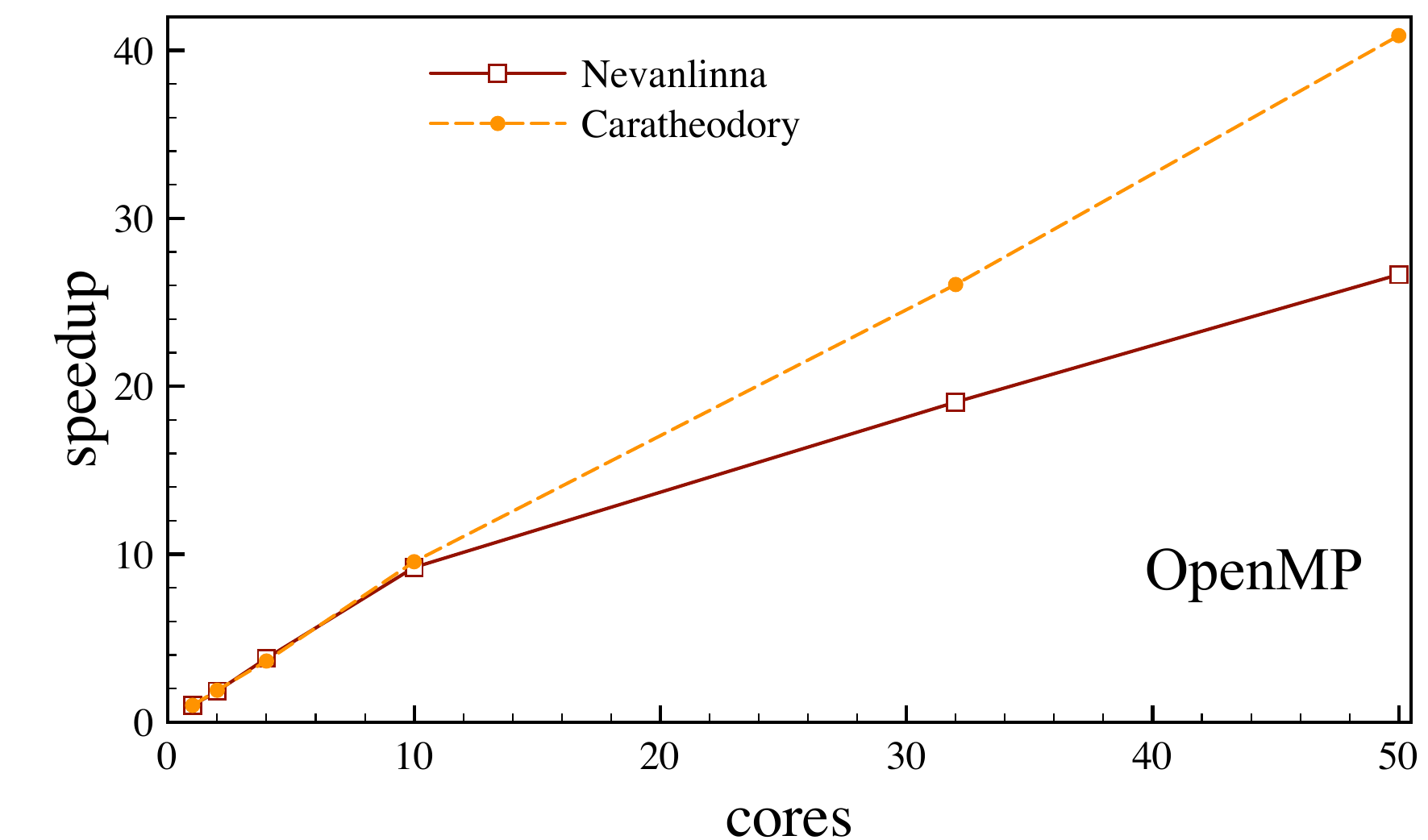}
    \includegraphics[width=.47\textwidth]{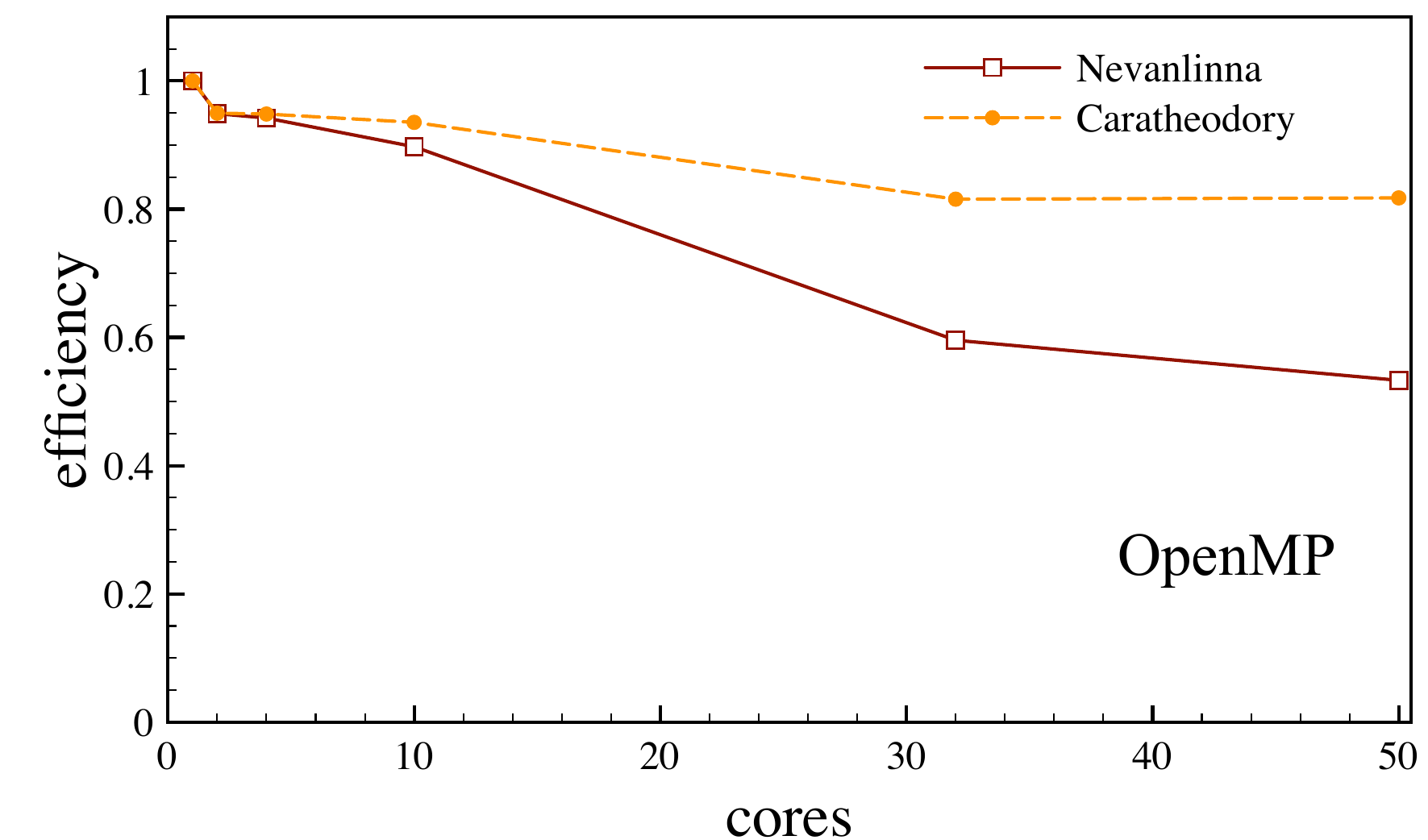}
    \caption{Parallel speedup (left) and efficiency (right) using MPI (top panel) and OpenMP (bottom panel)
    parallelization for Nevanlinna and Carath\'{e}odory continuation of 10 orbital Matsubara Green's function.}
    \label{fig:parallel_efficiency}
\end{figure}

\Nevanlinna is implemented with support of OpenMP and MPI parallelization. 
To choose number of OpenMP threads the environment variable \lstinline[language=Python]{NEVANLINNA_NUM_THREADS} 
is used. By default, the number of OpenMP threads are set to 1. When \Nevanlinna is run within MPI environment it
will be automatically detected and run in parallel. Fig.~\ref{fig:parallel_efficiency} shows parallel speedup and efficiency of MPI and OpenMP parallelization. In case of Carath\'{e}odory continuation, OpenMP scheme shows overall better
parallel performance due to larger computational needs for each loop iteration.

\subsection{Advanced usage}\label{subsec:adv}
\subsubsection{Finding smooth spectral functions}
As described in Ref.~\cite{Fei21},
the continued fraction form of the Nevanlinna factorization provides an interpolation of the input data irrespective of the choice of $\theta_{M+1}$, the last element in the factorization:
\begin{align}
    G(z) &= - h^{-1}\left[\theta(z)\right] \\
    \theta(z) &= \frac{a(z) \theta_{M+1}(z) +b(z)}{c(z)\theta_{M+1}(z) +d(z)},
\end{align}
where $a(z)$, $b(z)$, $c(z)$ and $d(z)$ are Nevanlinna factorization coefficients, and $h^{-1}$ is the inverse M\"obius transform.

For systems with discrete spectra, both Nevanlinna and Carath\'{e}odory continuation work well  irrespective of the choice of $\theta_{M+1}$   \cite{Fei21}.  $\theta_{M+1}$ is therefore usually set to zero. However, for systems with continuous spectral functions the continuation will likely return a highly oscillating function if $\theta_{M+1}=1$ . This is due to the fact that many spectral functions interpolate the data provided as input; all of these spectral functions can be expressed by adjusting $\theta_{M+1}$. The remaining freedom in the continuation problem can therefore be used to `smoothen' the resulting spectral function. This involves an expansion of the free Nevanlinna function into appropriate basis functions (we choose basis functions of a Hardy space), followed by the minimization of a function norm \cite{Fei21}.

This code implements a Hardy function optimization for Nevanlinna continuation. 
To perform the optimization, the user has to define a target function for the optimization problem that takes a retarded Green's function
as an input. The optimization is performed using SciPy's conjugate gradient method~\cite{SCIPY,Nocedal2006}. In the example below we show an
implementation for smoothness optimization by minimizing second derivative, as described in Ref.~\cite{Fei21}.

\begin{pylisting}
    w = np.array([x for x in m.values()])
    def SmoothnessOptimization (Gw):
        lmbda = 1e-4
        diff = 0.0
        # loop over all orbitals
        for i in range(Gw.data.shape[1]):
            # extract spectral function
            Aw = -Gw.data[:,i,i].imag/np.pi
            # compute estimate for second derivative
            Aw2 = (Aw[:-2] - 2*Aw[1:-1] + Aw[2:])/(Gw.mesh.delta**2)
            # evaluate normalization criteria weight
            diff += np.abs(1 - scipy.integrate.simpson(Aw, w))**2
            # evaluate smoothness criteria weight
            diff += lmbda * scipy.integrate.simpson(Aw2**2, w[1:-1])
        return diff

    # perform Hardy optimization
    G_w_opt = solver.optimize(m, eta, SmoothnessOptimization, gtol=1e-3, maxiter=1000)
\end{pylisting}

\subsubsection{Continuation of raw data}\label{subsubsec:raw}
In some cases a user may not have the option to provide \TRIQS Green's function objects to perform analytical continuation. For that
reason we provide direct access to continuation kernels classes and user simply can pass \verb#numpy# arrays as an input.
In the example below we show how to initialize kernel and use it on the raw data read from HDF5 file.

\begin{pylisting}
    # import numpy and h5py modules to prepare raw data for continuation
    import numpy as np
    import h5py
    # import continuation kernels
    from triqs_Nevanlinna import kernels

    # Initialize Nevanlinna kernel object
    kernel = kernels.NevanlinnaKernel()

    # Read raw data from HDF5 file
    with h5py.File("data.h5", "r") as input_data:
        data = input_data["g/iw/data"][()]
        mesh = input_data["g/iw/mesh"][()]

    # Make Matsubara frequencies purely imaginary
    mesh *= 1.j

    # Build Nevanlinna factorization
    kernel.init(mesh, data)

    # Define real frequency grid for continuation
    eta = 0.01
    w = np.linspace(-4, 4, 1000)
    w = w + 1.j*eta

    # Compute the diagonal part of the retarded Green's function
    Gw = kernel.evaluate(w)[:,:,:]

\end{pylisting}
Here the \lstinline[language=Python]{data} is a three-dimensional numpy array of size \lstinline[language=Python]{(niw,N,N)},
where $niw$ is the number of Matsubara frequencies and $N$ is the number of orbitals. Currently, we have two kernels available
\lstinline[language=Python]{NevanlinnaKernel} for scalar valued continuation and \lstinline[language=Python]{CaratheodoryKernel}
for matrix-valued continuation.

\subsubsection{C++ interface}
The C++ part of \Nevanlinna is designed as a library and uses CMake as a build system.
Adding it as a dependency into another CMake project is straightforward. First, one
 needs to include and link \Nevanlinna in the CMake project file
\begin{verbatim}
 find_package(triqs_Nevanlinna)
 target_link_libraries(TARGET_NAME triqs_Nevanlinna::triqs_Nevanlinna_c)
\end{verbatim}
Here \verb|TARGET_NAME| should be replaced according to the CMake target name in the project.
Second, the user needs to provide the path to the installed \Nevanlinna package by passing the following argument to CMake:
\begin{verbatim}
  -Dtriqs_Nevanlinna_DIR=<Nevanlinna_install_dir>/lib/cmake/triqs_Nevanlinna
\end{verbatim}

The example bellow shows a simple use case scenario reading imaginary time data from a HDF5 file into \TRIQS \verb|nda| arrays.
The file with input data and a source code of the example can also be found in the \verb|examples| directory of the github repository.

\begin{lstlisting}[language=C++]
#include <triqs_Nevanlinna/kernels.hpp>
#include <mpi/mpi.hpp>
#include <nda/nda.hpp>
#include <nda/h5.hpp>

using namespace std::literals;

int main(int argc, char **argv) {
  // Initialize MPI environment
  mpi::environment env(argc, argv);

  // Create kernel object
  triqs_Nevanlinna::Nevanlinna_kernel kernel;

  // Define real-frequency grid
  size_t N_w   = 5000;
  double w_min = -10., w_max = 10.;
  double eta = 0.1;
  auto del   = (w_max - w_min) / (N_w - 1);
  auto grid  = nda::basic_array{w_min + nda::arange(N_w) * del + eta * 1i};

  // Define imaginary time grid and data arrays
  auto mesh = nda::array<std::complex<double>, 1>{};
  auto G_iw = nda::array<std::complex<double>, 3>{};

  // Read imaginary time data from "input.h5" file
  auto input = h5::file(DATA_PATH + "/input.h5", 'r');
  h5::read(input, "data", G_iw);
  h5::read(input, "mesh", mesh);

  // Build Nevanlinna factorization
  kernel.init(mesh, G_iw);

  // Perform analytical continuation onto real frequency axis
  auto G_w = kernel.evaluate(grid);
}
\end{lstlisting}

\subsection{Analyzes and Troubleshooting}\label{subsec:analyzes}
Both Nevanlinna and Carath\'{e}odory continuations interpolate, rather than fit, Matsubara data and are therefore susceptible to the quality of the input data.
The existence of a causal continuation can be guaranteed only if the input data corresponds to a positive semi-definite matrix-valued function anywhere in the upper-half of the complex plane. 
In numerical double precision, this is almost never the case due to numerical round-off. Nevertheless, a simple check on the quality of the input data is given by the eigenvalues of the Pick matrix of the input data \cite{Pick17}. 
For precise causal data, negative eigenvalues of the Pick matrix will lie within the numerical diagonalization threshold. To obtain eigenvalues of the Pick matrix in this code, a user may read the \lstinline[language=Python]{get_Pick_eigenvalues} property
from the solver object. Fig.~\ref{fig:Pick_eigenvalues} shows typical behavior of Pick's matrix eigenvalues obtained for
data without numerical noise (left panel) and for data with numerical noise (right panel). Since the largest negative eigenvalue $|\nu| \gg 0$,
no causal interpolation of the input data exists, and therefore a successful analytical continuation cannot be guaranteed.

\begin{figure}
    \includegraphics[width=.47\textwidth]{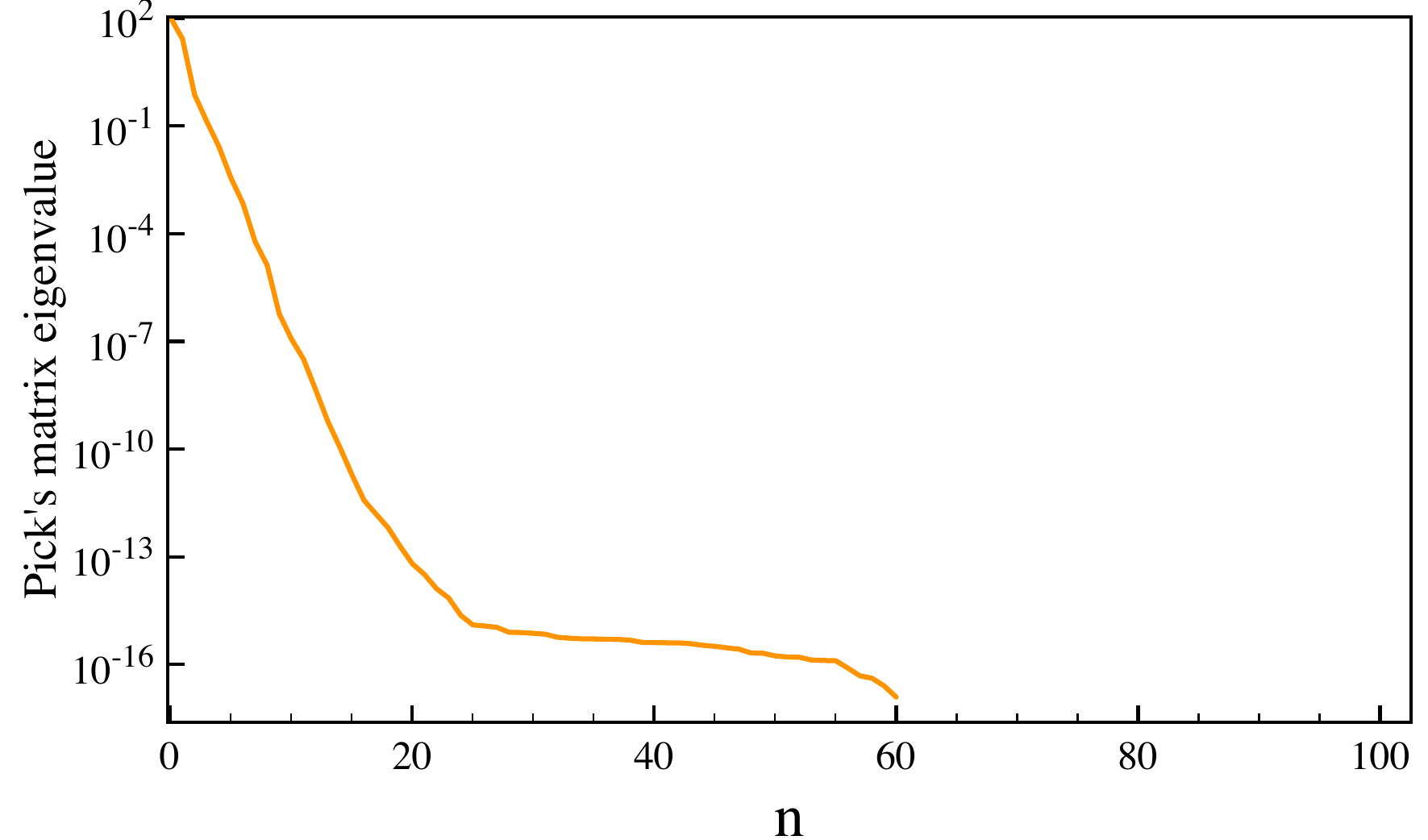}
    \includegraphics[width=.47\textwidth]{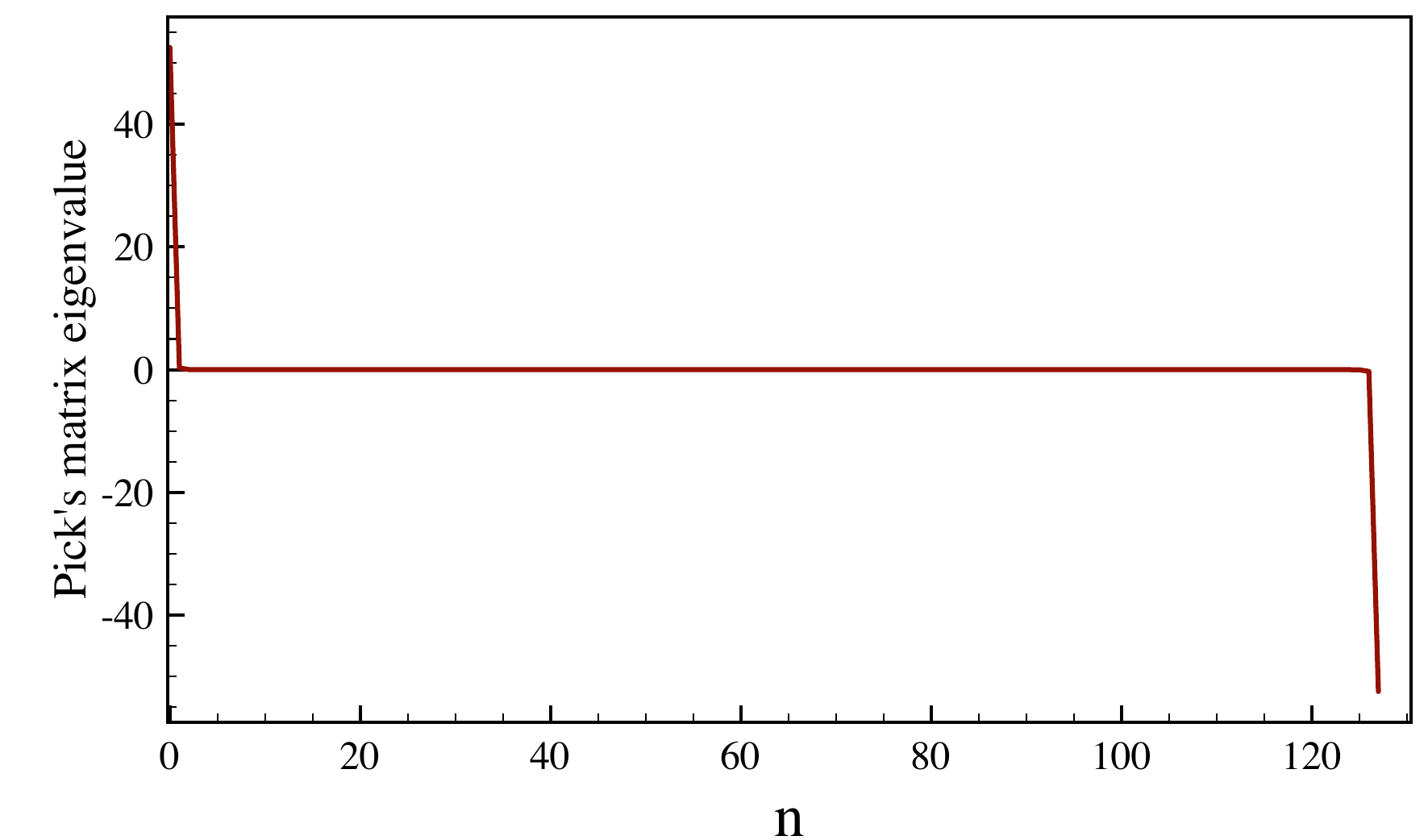}
    \caption{Eigenvalues of the Pick matrix for data obtain from solution of a multi-orbital impurity problem with Exact Diagonalization (left)
        and Continous-Time QMC (right).}
    \label{fig:Pick_eigenvalues}
\end{figure}

\section{Examples}\label{sec:examples}
We demonstrate the application for scalar-valued functions using the Nevanlinna formalism.

\subsection{Scalar-valued function continuation}\label{subsec:scalar-valued}
We consider four exactly solvable examples: the non-interacting Bethe lattice, the non-interacting square lattice, the Hubbard atom, and
the single impurity Anderson model. For the Hardy functions optimization we choose the smoothness criteria as described in the Ref.~\cite{Fei21}.

The first model we consider is non-interacting Bethe lattice, with
\begin{align}
    G(z) = \frac{z - \sqrt{4 t^2 - z^2}}{2 t^2},
\end{align}
with general complex frequency $z$ and hopping $t$. For Matsubara frequencies $z = i \omega_n$ and for real frequencies $z= \omega+ i\delta$,
where $\delta$ is Lorentzian broadening parameter. We choose $\delta=0.1$ for all examples. The left panel in Fig.~\ref{fig:model_examples} shows the resulting unoptimized spectral
function (dashed red line) and its comparison to the exact semicircular density of states (black line). As expected the unoptimized spectral
function consists of a set of sharp peaks. After performing the Hardy functions optimization (see Sec.\ref{subsec:adv} for details) we are able to recover smoothness of the spectral function away from the boundaries.

Similarly to Bethe lattice, we show results for a non-interacting square lattice local density of states in the right panel of Fig.~\ref{fig:model_examples}.
Again, the Hardy function optimization allows to substantially suppress sharp features in the continuation. It is worth to mention that in
both these examples second derivative does not exist at the spectral boundaries and possibly different type of optimization could be
necessary.

\begin{figure}
    \includegraphics[width=.47\textwidth]{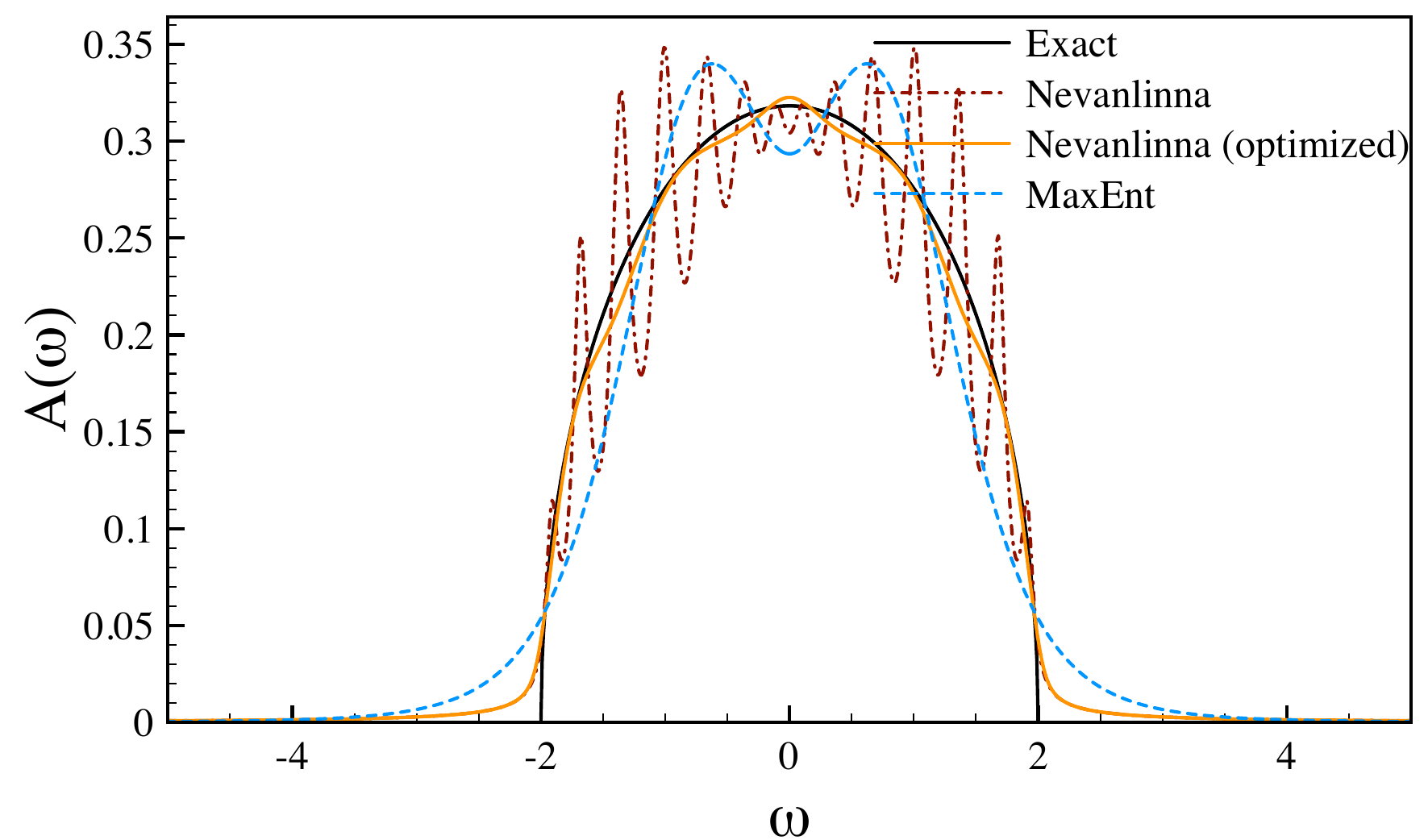}
    \includegraphics[width=.47\textwidth]{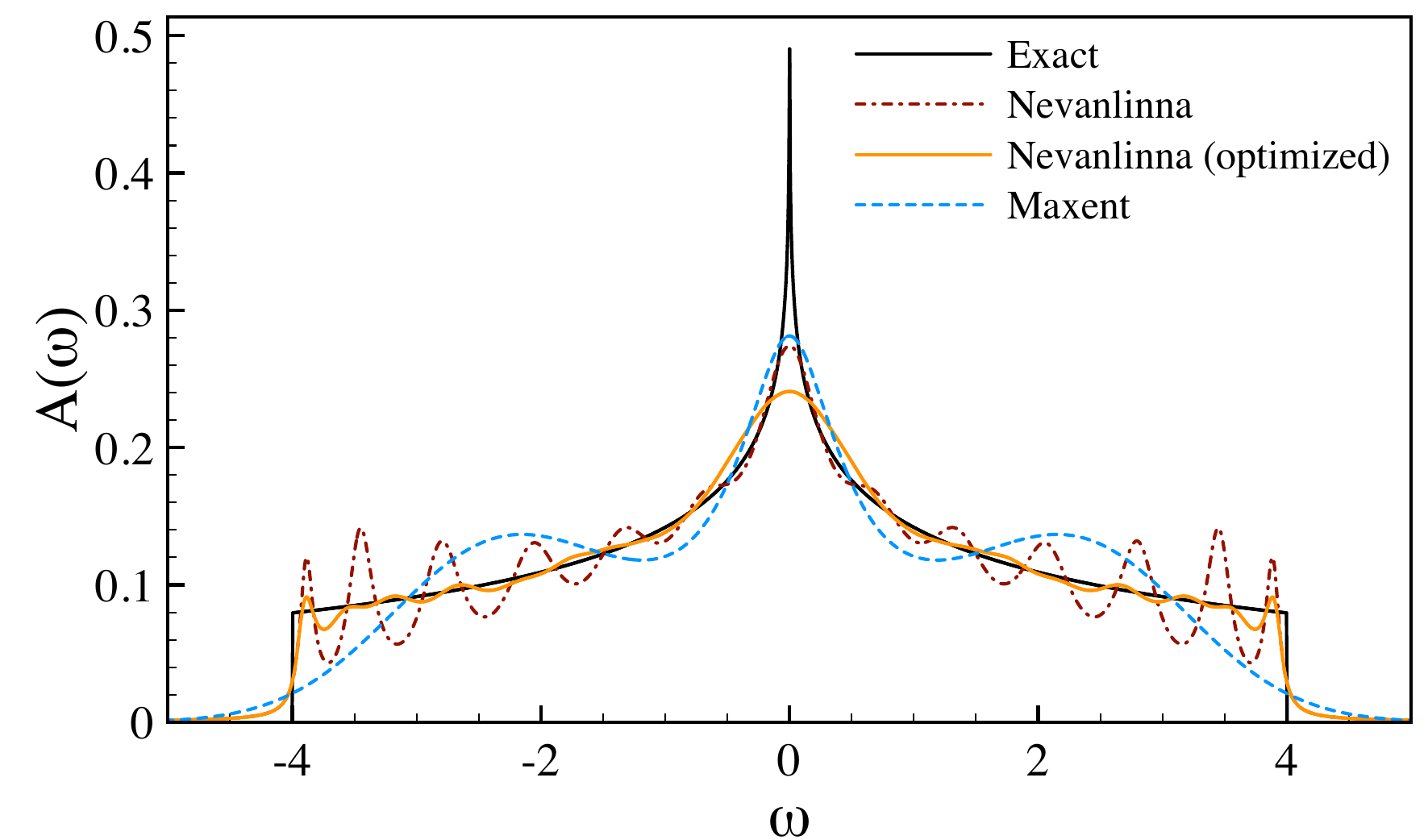}
    \caption{Continuation results for analytically solvable models. Left panel: Semi-circular density of states for non-interacting Bethe lattice.
    Right panel: non-interacting square lattice density of states.
    Black lines: exact solution.
    Dashed red: Nevanlinna continuation.
    Orange: Nevanlinna continuation with Hardy function optimization.
    Dashed blue: Maxent analytical continuation.}
    \label{fig:model_examples} 
\end{figure}

\begin{figure}
\includegraphics[width=.47\textwidth]{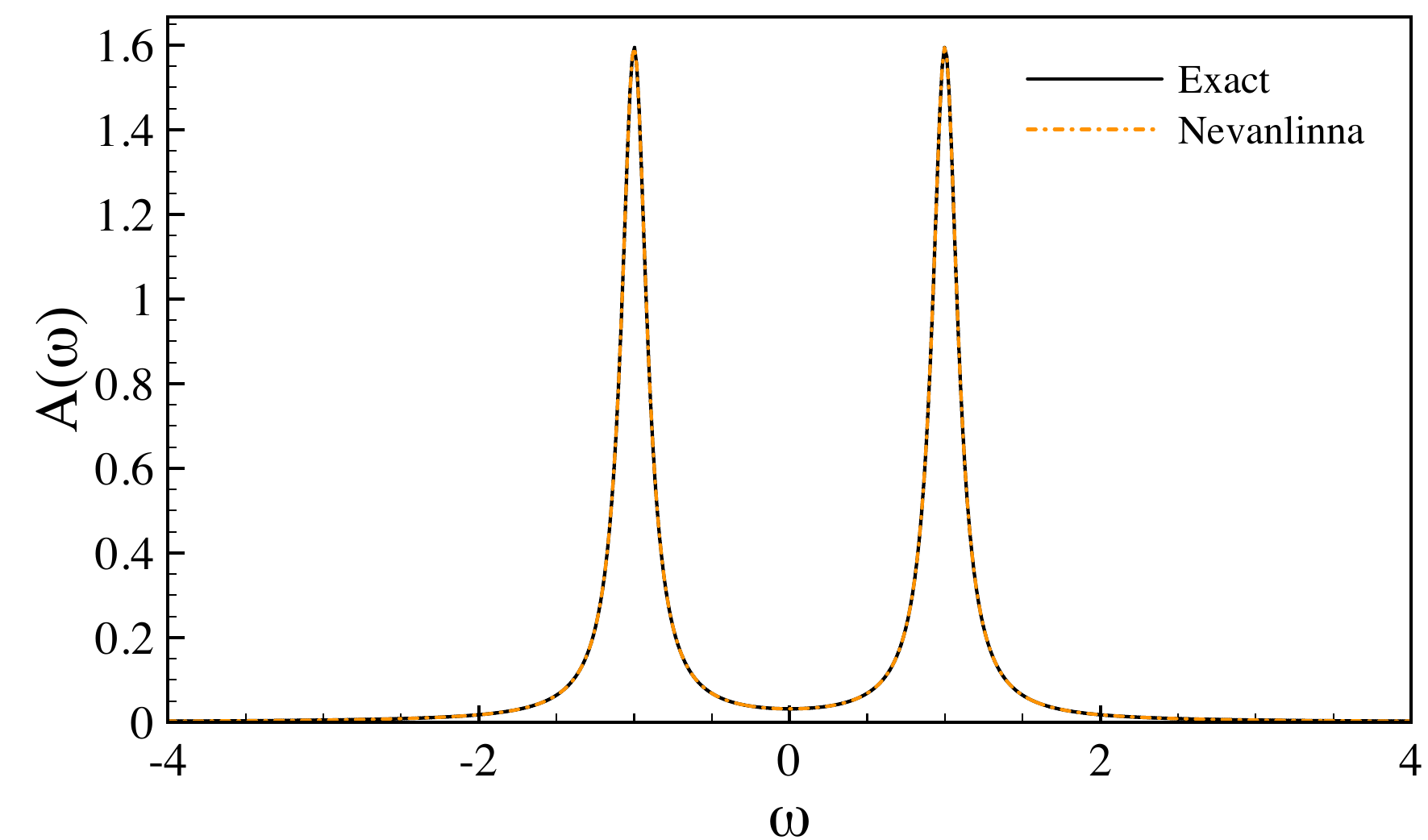}
\includegraphics[width=.47\textwidth]{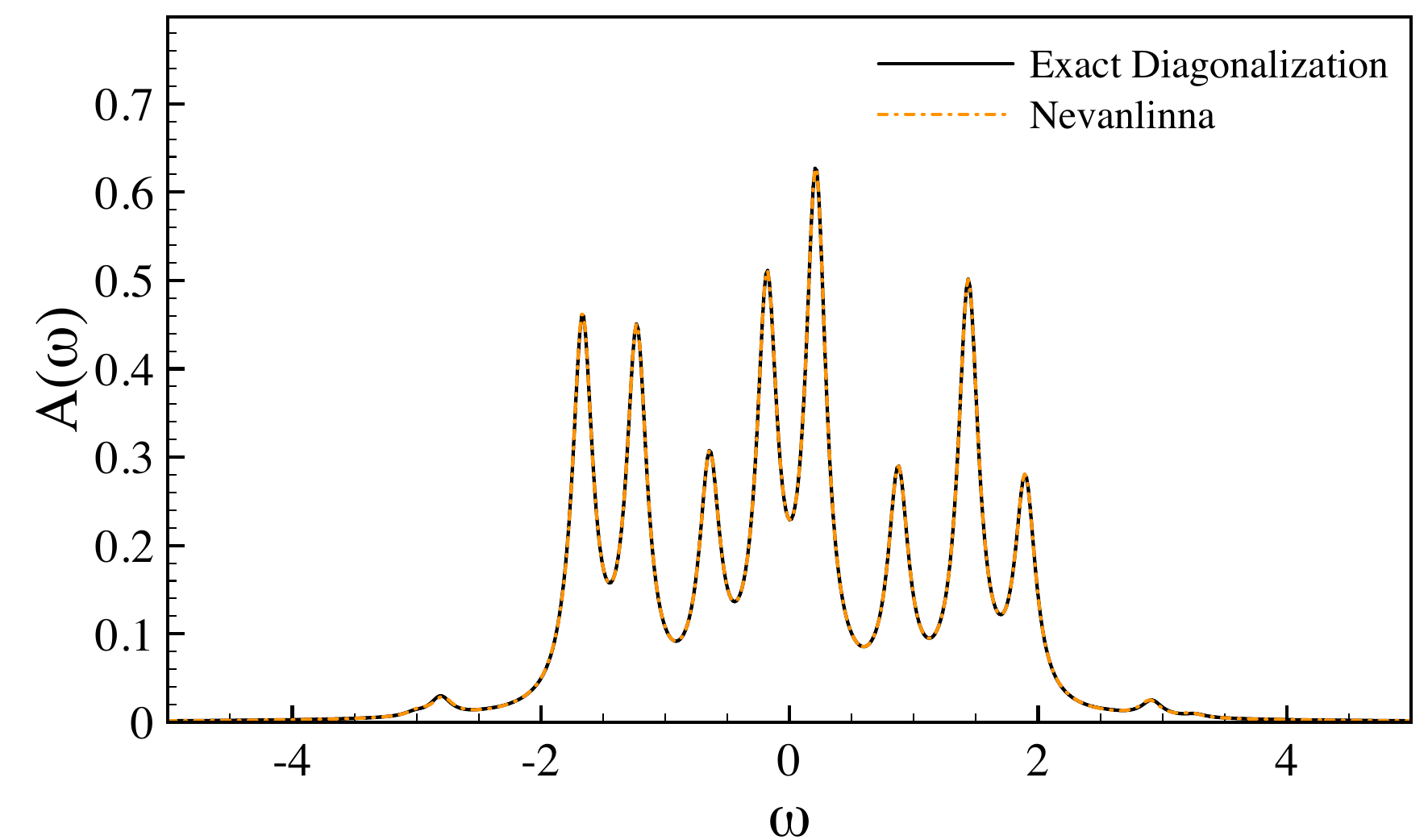}
\caption{
Results for analytical continuation of exactly solvable models. Left panel: half-filled Hubbard atom at $U=2.0$. Right panel:
single orbital Anderson impurity model for $\mu = 1.0$, $U = 2.0$, with bath parameters $E = \left[ -0.5, 0.7 \right]$ and $V = \left[ 0.4, 0.5 \right]$.
Black lines: exact diagonalization, dashed orange lines: Nevanlinna analytical continuation.
}
\label{fig:model_examples2}
\end{figure}

Next we consider the single Hubbard atom with the Coulomb interaction $U = 2$ at half-filling. Its spectral function
has discrete levels at $\omega=\pm\frac{U}{2}$. The Green’s function for general complex frequency of the Hubbard atom is
\begin{align}
    G(z) = \frac{1}{2}\left[\frac{1}{z + \frac{U}{2}} + \frac{1}{z - \frac{U}{2}}\right]
\end{align}
The left panel of Fig.~\ref{fig:model_examples2} shows Nevanlinna continuation (dashed orange) and exact (black)
results. As expected Nevanlinna continuation perfectly resolved sharp peaks for that model.

The last example shows results for the discrete single impurity Anderson model that is described by the Hamiltonian
\begin{align}
H &=  -\mu \, (n_{\uparrow, 0} + n_{\uparrow, 0}) - h(n_{\uparrow,0} - n_{\downarrow,0})  + U n_{\uparrow, 0} n_{\downarrow, 0}\nonumber \\
&+\sum_{i=1}^{N} \sum_{\sigma} E_i n_{\sigma,i} + \sum_{i=1}^{N} \sum_{\sigma} V_i (c^\dagger_{\sigma,i} c_{\sigma,0} + c^\dagger_{\sigma,0} c_{\sigma,i}),
\end{align}
where $\mu$ is the chemical potential, $U$ is the local interaction, $E_i$ is the bath electrons energy and $V_i$ is the coupling between impurity and bath electrons.
This model does not have a simple closed-form solution and we use exact diagonalization to obtain both the Matsubara and real frequency Green's function using the Lehmann representation for a small number of bath sites $N=2$.
We choose $\mu = 1.0$, $U = 2.0$, $E = \left[ -0.5, 0.7 \right]$ and $V = \left[ 0.4, 0.5 \right]$. The right panel of Fig.~\ref{fig:model_examples2}
shows the results from exact diagonalization (black) and Nevanlinna continuation (dashed orange). Similarly to the Hubbard atom,
Nevanlinna continuation perfectly resolves all the poles in the real frequency Green's function.

\subsection{Matrix-valued function continuation}\label{subsec:matrix-valued}

For the matrix valued continuation we consider the discrete two orbital Anderson impurity model with inter-orbital hopping that is described by the following Hamiltonian
\begin{align}
    H &=  -\mu \sum_{i=0,1} (n_{\uparrow, i} + n_{\uparrow, i}) - h(n_{\uparrow,0} - n_{\downarrow,0})  + U n_{\uparrow, 0} n_{\downarrow, 0}\nonumber \\
    &+\sum_{\sigma} t_{ij}(c^\dagger_{\sigma,0} c_{\sigma,1} + c^\dagger_{\sigma,1} c_{\sigma,0}) + V \frac{1}{2}\sum_{\sigma\sigma'}(n_{\sigma, 0} n_{\sigma', 1}) \nonumber\\
    &+\sum_{k=2}^{N} \sum_{\sigma} E_k n_{\sigma,k} + \sum_{i=0,1}\sum_{k=2}^{N} \sum_{\sigma} V_k (c^\dagger_{\sigma,k} c_{\sigma,i} + c^\dagger_{\sigma,i} c_{\sigma,k}),
    \label{eqn:SIAM_two}
\end{align}
with inter-orbital hopping $t$ and inter-orbital interaction $V$, for two bath sites $N=2$.

\begin{figure}
    \includegraphics[width=.47\textwidth]{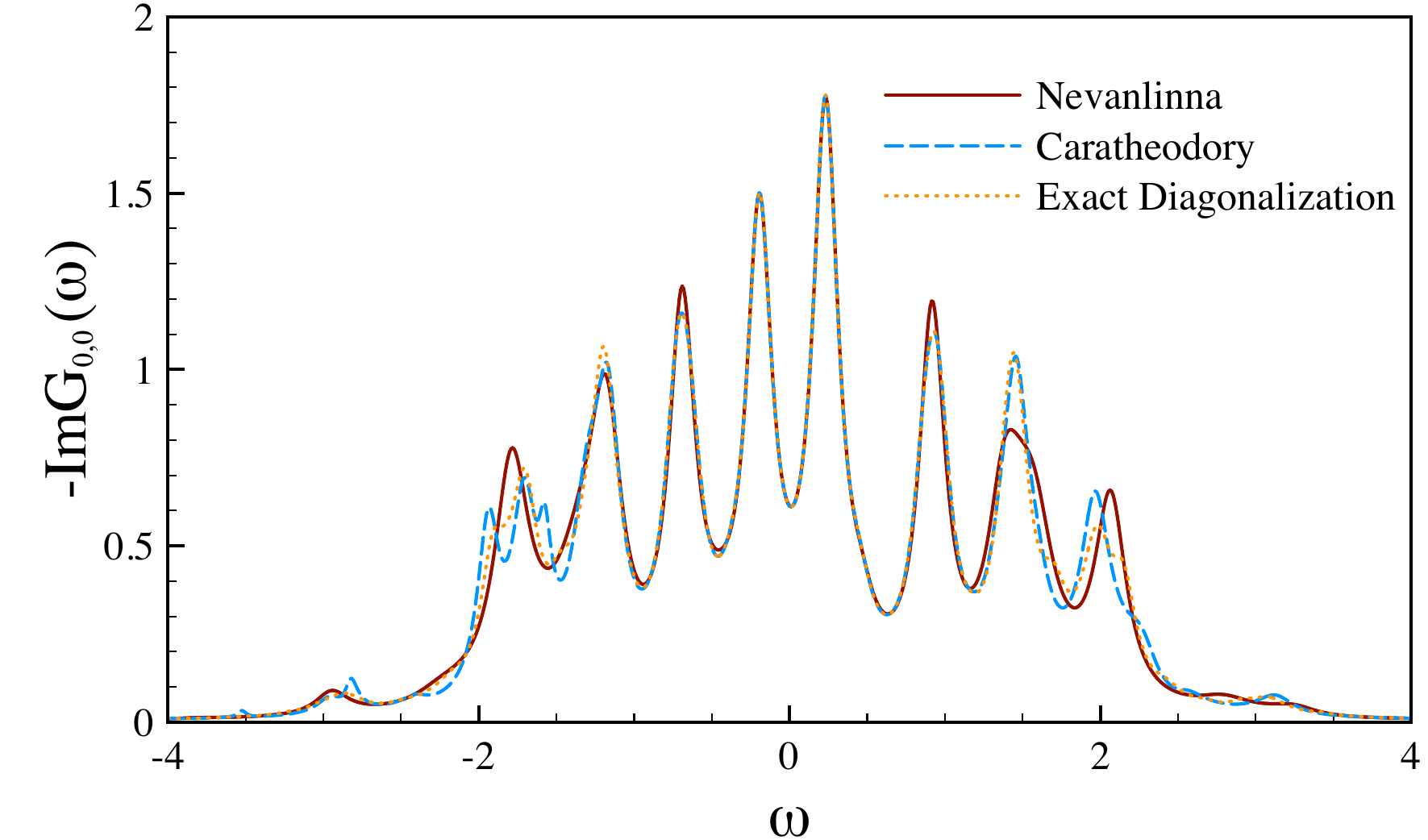}
    \includegraphics[width=.47\textwidth]{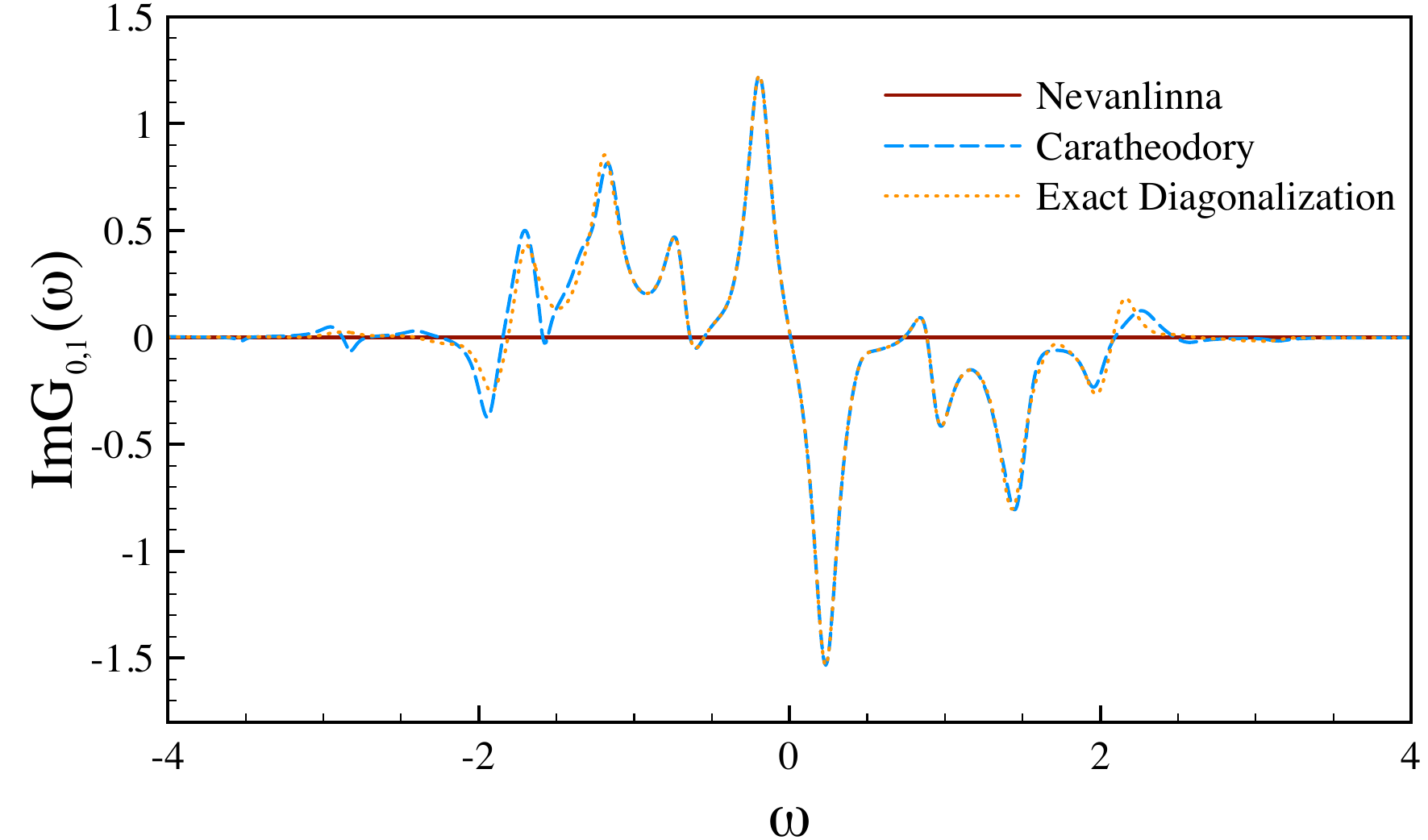}
    \caption{
    Analytical continuation of the exact solution of two-orbital Anderson model, left panel: diagonal part of the Green's function,
        right panel: inter-orbital part of the Green's function. Dark red: Nevanlinna analytical continuation.
    Dashed blue: Carath\'{e}odory matrix-valued analytical continuation.
    Dotted: Exact diagonalization reference solution.
    }
    \label{fig:model_examples_mat}
\end{figure}

Fig.~\ref{fig:model_examples_mat} shows the results for analytical continuation results of the two-orbital Anderson impurity model (Eq.~\ref{eqn:SIAM_two})
for $U=2.0$, $V=0.1$, $\mu = 1.0$, $t=0.3$, and bath parameters $E_k = [ -0.5, 0.7 ]$ and $V_k = [ 0.4, 0.5 ]$, at inverse temperature $\beta=40$ and brodening
parameter $\eta = 0.1$.
Dark red lines show Nevanlinna continuation and dashed blue lines correspond to Carath\'{e}odory continuation. As the Nevanlinna continuation kernel
only allows to continue the diagonal part (left panel), the off-diagonal part (right panel) is only relevant when using the Carath\'{e}odory continuation.

\section{Summary}\label{sec:summary}
We have presented the open-source \TRIQS/\Nevanlinna package that implements the analytic continuation method for noise free data providing
two kernels for scalar-valued and matrix-valued function.
We demonstrated the effectiveness of the \TRIQS/\Nevanlinna implementation for obtaining spectral functions for both model systems
and realistic materials. We also provide Hardy function optimization for scalar-valued functions to optimize the continuation of continuos spectra. The current implementation allows user to pass a target function to perform desired optimization.

Future releases will include additional kernels for noisy data and continuation optimization for matrix-valued functions. The presented application is based upon the TRIQS software library and is part of its ecosystem, ensuring support for new compilers and libraries, and flexibility when used in other applications.

\section{Acknowledgments}
S.I. was supported by the Simons Foundation. E.G. was supported by the National Science Foundation under Grant No. 2001465. The Flatiron Institute is a division of the Simons Foundation.

\bibliographystyle{elsarticle-num.bst}
\bibliography{main}

\begin{thebibliography}{10}
\expandafter\ifx\csname url\endcsname\relax
  \def\url#1{\texttt{#1}}\fi
\expandafter\ifx\csname urlprefix\endcsname\relax\def\urlprefix{URL }\fi
\expandafter\ifx\csname href\endcsname\relax
  \def\href#1#2{#2} \def\path#1{#1}\fi

\bibitem{TRIQS}
O.~Parcollet, M.~Ferrero, T.~Ayral, H.~Hafermann, I.~Krivenko, L.~Messio,
  P.~Seth, Triqs: A toolbox for research on interacting quantum systems,
  Computer Physics Communications 196 (2015) 398--415.
\newblock \href {https://doi.org/https://doi.org/10.1016/j.cpc.2015.04.023}
  {\path{doi:https://doi.org/10.1016/j.cpc.2015.04.023}}.

\bibitem{Mahan00}
G.~Mahan, \href{https://books.google.com/books?id=xzSgZ4-yyMEC}{Many-Particle
  Physics}, Physics of Solids and Liquids, Springer, 2000.
\newline\urlprefix\url{https://books.google.com/books?id=xzSgZ4-yyMEC}

\bibitem{Ceperley95}
D.~M. Ceperley, \href{https://link.aps.org/doi/10.1103/RevModPhys.67.279}{Path
  integrals in the theory of condensed helium}, Rev. Mod. Phys. 67 (1995)
  279--355.
\newblock \href {https://doi.org/10.1103/RevModPhys.67.279}
  {\path{doi:10.1103/RevModPhys.67.279}}.
\newline\urlprefix\url{https://link.aps.org/doi/10.1103/RevModPhys.67.279}

\bibitem{Blankenbecler81}
R.~Blankenbecler, D.~J. Scalapino, R.~L. Sugar,
  \href{https://link.aps.org/doi/10.1103/PhysRevD.24.2278}{Monte carlo
  calculations of coupled boson-fermion systems. i}, Phys. Rev. D 24 (1981)
  2278--2286.
\newblock \href {https://doi.org/10.1103/PhysRevD.24.2278}
  {\path{doi:10.1103/PhysRevD.24.2278}}.
\newline\urlprefix\url{https://link.aps.org/doi/10.1103/PhysRevD.24.2278}

\bibitem{Scalapino81}
D.~J. Scalapino, R.~L. Sugar,
  \href{https://link.aps.org/doi/10.1103/PhysRevB.24.4295}{Monte carlo
  calculations of coupled boson-fermion systems. ii}, Phys. Rev. B 24 (1981)
  4295--4308.
\newblock \href {https://doi.org/10.1103/PhysRevB.24.4295}
  {\path{doi:10.1103/PhysRevB.24.4295}}.
\newline\urlprefix\url{https://link.aps.org/doi/10.1103/PhysRevB.24.4295}

\bibitem{Hedin65}
L.~Hedin, \href{https://link.aps.org/doi/10.1103/PhysRev.139.A796}{New method
  for calculating the one-particle green's function with application to the
  electron-gas problem}, Phys. Rev. 139 (1965) A796--A823.
\newblock \href {https://doi.org/10.1103/PhysRev.139.A796}
  {\path{doi:10.1103/PhysRev.139.A796}}.
\newline\urlprefix\url{https://link.aps.org/doi/10.1103/PhysRev.139.A796}

\bibitem{Holleboom90}
L.~J. Holleboom, J.~G. Snijders, \href{https://doi.org/10.1063/1.459578}{{A
  comparison between the Mo/ller–Plesset and Green’s function perturbative
  approaches to the calculation of the correlation energy in the
  many‐electron problem}}, The Journal of Chemical Physics 93~(8) (1990)
  5826--5837.
\newblock \href
  {http://arxiv.org/abs/https://pubs.aip.org/aip/jcp/article-pdf/93/8/5826/11207185/5826\_1\_online.pdf}
  {\path{arXiv:https://pubs.aip.org/aip/jcp/article-pdf/93/8/5826/11207185/5826\_1\_online.pdf}},
  \href {https://doi.org/10.1063/1.459578} {\path{doi:10.1063/1.459578}}.
\newline\urlprefix\url{https://doi.org/10.1063/1.459578}

\bibitem{Dahlen06}
N.~E. Dahlen, R.~van Leeuwen, U.~von Barth,
  \href{https://link.aps.org/doi/10.1103/PhysRevA.73.012511}{Variational energy
  functionals of the green function and of the density tested on molecules},
  Phys. Rev. A 73 (2006) 012511.
\newblock \href {https://doi.org/10.1103/PhysRevA.73.012511}
  {\path{doi:10.1103/PhysRevA.73.012511}}.
\newline\urlprefix\url{https://link.aps.org/doi/10.1103/PhysRevA.73.012511}

\bibitem{Stan09}
A.~Stan, N.~E. Dahlen, R.~van Leeuwen,
  \href{https://doi.org/10.1063/1.3089567}{{Levels of self-consistency in the
  GW approximation}}, The Journal of Chemical Physics 130~(11) (2009) 114105.
\newblock \href
  {http://arxiv.org/abs/https://pubs.aip.org/aip/jcp/article-pdf/doi/10.1063/1.3089567/15424617/114105\_1\_online.pdf}
  {\path{arXiv:https://pubs.aip.org/aip/jcp/article-pdf/doi/10.1063/1.3089567/15424617/114105\_1\_online.pdf}},
  \href {https://doi.org/10.1063/1.3089567} {\path{doi:10.1063/1.3089567}}.
\newline\urlprefix\url{https://doi.org/10.1063/1.3089567}

\bibitem{Phillips14}
J.~J. Phillips, D.~Zgid,
  \href{https://doi.org/10.1063/1.4884951}{{Communication: The description of
  strong correlation within self-consistent Green's function second-order
  perturbation theory}}, The Journal of Chemical Physics 140~(24) (2014)
  241101.
\newblock \href
  {http://arxiv.org/abs/https://pubs.aip.org/aip/jcp/article-pdf/doi/10.1063/1.4884951/14082240/241101\_1\_online.pdf}
  {\path{arXiv:https://pubs.aip.org/aip/jcp/article-pdf/doi/10.1063/1.4884951/14082240/241101\_1\_online.pdf}},
  \href {https://doi.org/10.1063/1.4884951} {\path{doi:10.1063/1.4884951}}.
\newline\urlprefix\url{https://doi.org/10.1063/1.4884951}

\bibitem{Rusakov16}
A.~A. Rusakov, D.~Zgid, Self-consistent second-order green's function
  perturbation theory for periodic systems., J Chem Phys 144~(5) (2016) 054106.
\newblock \href {https://doi.org/10.1063/1.4940900}
  {\path{doi:10.1063/1.4940900}}.

\bibitem{Yeh22}
C.-N. Yeh, S.~Iskakov, D.~Zgid, E.~Gull,
  \href{https://link.aps.org/doi/10.1103/PhysRevB.106.235104}{Fully
  self-consistent finite-temperature $gw$ in gaussian bloch orbitals for
  solids}, Phys. Rev. B 106 (2022) 235104.
\newblock \href {https://doi.org/10.1103/PhysRevB.106.235104}
  {\path{doi:10.1103/PhysRevB.106.235104}}.
\newline\urlprefix\url{https://link.aps.org/doi/10.1103/PhysRevB.106.235104}

\bibitem{Metzner89}
W.~Metzner, D.~Vollhardt,
  \href{https://link.aps.org/doi/10.1103/PhysRevLett.62.324}{Correlated lattice
  fermions in $d=\ensuremath{\infty}$ dimensions}, Phys. Rev. Lett. 62 (1989)
  324--327.
\newblock \href {https://doi.org/10.1103/PhysRevLett.62.324}
  {\path{doi:10.1103/PhysRevLett.62.324}}.
\newline\urlprefix\url{https://link.aps.org/doi/10.1103/PhysRevLett.62.324}

\bibitem{Georges92}
A.~Georges, G.~Kotliar,
  \href{https://link.aps.org/doi/10.1103/PhysRevB.45.6479}{Hubbard model in
  infinite dimensions}, Phys. Rev. B 45 (1992) 6479--6483.
\newblock \href {https://doi.org/10.1103/PhysRevB.45.6479}
  {\path{doi:10.1103/PhysRevB.45.6479}}.
\newline\urlprefix\url{https://link.aps.org/doi/10.1103/PhysRevB.45.6479}

\bibitem{Georges96}
A.~Georges, G.~Kotliar, W.~Krauth, M.~J. Rozenberg,
  \href{https://link.aps.org/doi/10.1103/RevModPhys.68.13}{Dynamical mean-field
  theory of strongly correlated fermion systems and the limit of infinite
  dimensions}, Rev. Mod. Phys. 68 (1996) 13--125.
\newblock \href {https://doi.org/10.1103/RevModPhys.68.13}
  {\path{doi:10.1103/RevModPhys.68.13}}.
\newline\urlprefix\url{https://link.aps.org/doi/10.1103/RevModPhys.68.13}

\bibitem{Kotliar06}
G.~Kotliar, S.~Y. Savrasov, K.~Haule, V.~S. Oudovenko, O.~Parcollet, C.~A.
  Marianetti,
  \href{https://link.aps.org/doi/10.1103/RevModPhys.78.865}{Electronic
  structure calculations with dynamical mean-field theory}, Rev. Mod. Phys. 78
  (2006) 865--951.
\newblock \href {https://doi.org/10.1103/RevModPhys.78.865}
  {\path{doi:10.1103/RevModPhys.78.865}}.
\newline\urlprefix\url{https://link.aps.org/doi/10.1103/RevModPhys.78.865}

\bibitem{Zgid17}
D.~Zgid, E.~Gull, \href{https://dx.doi.org/10.1088/1367-2630/aa5d34}{Finite
  temperature quantum embedding theories for correlated systems}, New Journal
  of Physics 19~(2) (2017) 023047.
\newblock \href {https://doi.org/10.1088/1367-2630/aa5d34}
  {\path{doi:10.1088/1367-2630/aa5d34}}.
\newline\urlprefix\url{https://dx.doi.org/10.1088/1367-2630/aa5d34}

\bibitem{Rusakov19}
A.~A. Rusakov, S.~Iskakov, L.~N. Tran, D.~Zgid,
  \href{https://doi.org/10.1021/acs.jctc.8b00927}{Self-energy embedding theory
  (seet) for periodic systems}, Journal of Chemical Theory and Computation
  15~(1) (2019) 229--240, pMID: 30540474.
\newblock \href {http://arxiv.org/abs/https://doi.org/10.1021/acs.jctc.8b00927}
  {\path{arXiv:https://doi.org/10.1021/acs.jctc.8b00927}}, \href
  {https://doi.org/10.1021/acs.jctc.8b00927}
  {\path{doi:10.1021/acs.jctc.8b00927}}.
\newline\urlprefix\url{https://doi.org/10.1021/acs.jctc.8b00927}

\bibitem{Prokofev98}
N.~V. Prokof'ev, B.~V. Svistunov,
  \href{https://link.aps.org/doi/10.1103/PhysRevLett.81.2514}{Polaron problem
  by diagrammatic quantum monte carlo}, Phys. Rev. Lett. 81 (1998) 2514--2517.
\newblock \href {https://doi.org/10.1103/PhysRevLett.81.2514}
  {\path{doi:10.1103/PhysRevLett.81.2514}}.
\newline\urlprefix\url{https://link.aps.org/doi/10.1103/PhysRevLett.81.2514}

\bibitem{Prokofev07}
N.~Prokof'ev, B.~Svistunov,
  \href{https://link.aps.org/doi/10.1103/PhysRevLett.99.250201}{Bold
  diagrammatic monte carlo technique: When the sign problem is welcome}, Phys.
  Rev. Lett. 99 (2007) 250201.
\newblock \href {https://doi.org/10.1103/PhysRevLett.99.250201}
  {\path{doi:10.1103/PhysRevLett.99.250201}}.
\newline\urlprefix\url{https://link.aps.org/doi/10.1103/PhysRevLett.99.250201}

\bibitem{Rubtsov05}
A.~N. Rubtsov, V.~V. Savkin, A.~I. Lichtenstein,
  \href{https://link.aps.org/doi/10.1103/PhysRevB.72.035122}{Continuous-time
  quantum monte carlo method for fermions}, Phys. Rev. B 72 (2005) 035122.
\newblock \href {https://doi.org/10.1103/PhysRevB.72.035122}
  {\path{doi:10.1103/PhysRevB.72.035122}}.
\newline\urlprefix\url{https://link.aps.org/doi/10.1103/PhysRevB.72.035122}

\bibitem{Werner06}
P.~Werner, A.~Comanac, L.~de' Medici, M.~Troyer, A.~J. Millis,
  \href{https://link.aps.org/doi/10.1103/PhysRevLett.97.076405}{Continuous-time
  solver for quantum impurity models}, Phys. Rev. Lett. 97 (2006) 076405.
\newblock \href {https://doi.org/10.1103/PhysRevLett.97.076405}
  {\path{doi:10.1103/PhysRevLett.97.076405}}.
\newline\urlprefix\url{https://link.aps.org/doi/10.1103/PhysRevLett.97.076405}

\bibitem{Gull08}
E.~Gull, P.~Werner, O.~Parcollet, M.~Troyer,
  \href{https://dx.doi.org/10.1209/0295-5075/82/57003}{Continuous-time
  auxiliary-field monte carlo for quantum impurity models}, Europhysics Letters
  82~(5) (2008) 57003.
\newblock \href {https://doi.org/10.1209/0295-5075/82/57003}
  {\path{doi:10.1209/0295-5075/82/57003}}.
\newline\urlprefix\url{https://dx.doi.org/10.1209/0295-5075/82/57003}

\bibitem{Gull11}
E.~Gull, A.~J. Millis, A.~I. Lichtenstein, A.~N. Rubtsov, M.~Troyer, P.~Werner,
  \href{https://link.aps.org/doi/10.1103/RevModPhys.83.349}{Continuous-time
  monte carlo methods for quantum impurity models}, Rev. Mod. Phys. 83 (2011)
  349--404.
\newblock \href {https://doi.org/10.1103/RevModPhys.83.349}
  {\path{doi:10.1103/RevModPhys.83.349}}.
\newline\urlprefix\url{https://link.aps.org/doi/10.1103/RevModPhys.83.349}

\bibitem{Jarrell96}
M.~Jarrell, J.~Gubernatis,
  \href{https://www.sciencedirect.com/science/article/pii/0370157395000747}{Bayesian
  inference and the analytic continuation of imaginary-time quantum monte carlo
  data}, Physics Reports 269~(3) (1996) 133--195.
\newblock \href {https://doi.org/https://doi.org/10.1016/0370-1573(95)00074-7}
  {\path{doi:https://doi.org/10.1016/0370-1573(95)00074-7}}.
\newline\urlprefix\url{https://www.sciencedirect.com/science/article/pii/0370157395000747}

\bibitem{Bergeron16}
D.~Bergeron, A.-M.~S. Tremblay,
  \href{https://link.aps.org/doi/10.1103/PhysRevE.94.023303}{Algorithms for
  optimized maximum entropy and diagnostic tools for analytic continuation},
  Phys. Rev. E 94 (2016) 023303.
\newblock \href {https://doi.org/10.1103/PhysRevE.94.023303}
  {\path{doi:10.1103/PhysRevE.94.023303}}.
\newline\urlprefix\url{https://link.aps.org/doi/10.1103/PhysRevE.94.023303}

\bibitem{Levy17}
R.~Levy, J.~LeBlanc, E.~Gull,
  \href{https://www.sciencedirect.com/science/article/pii/S0010465517300309}{Implementation
  of the maximum entropy method for analytic continuation}, Computer Physics
  Communications 215 (2017) 149--155.
\newblock \href {https://doi.org/https://doi.org/10.1016/j.cpc.2017.01.018}
  {\path{doi:https://doi.org/10.1016/j.cpc.2017.01.018}}.
\newline\urlprefix\url{https://www.sciencedirect.com/science/article/pii/S0010465517300309}

\bibitem{Huang22}
L.~Huang, Acflow: An open source toolkit for analytical continuation of quantum
  monte carlo data, SSRN Electronic Journal (2022).

\bibitem{Kaufmann23}
J.~Kaufmann, K.~Held,
  \href{https://www.sciencedirect.com/science/article/pii/S0010465522002387}{anacont
  python package for analytic continuation}, Computer Physics Communications
  282 (2023) 108519.
\newblock \href {https://doi.org/https://doi.org/10.1016/j.cpc.2022.108519}
  {\path{doi:https://doi.org/10.1016/j.cpc.2022.108519}}.
\newline\urlprefix\url{https://www.sciencedirect.com/science/article/pii/S0010465522002387}

\bibitem{Burnier13}
Y.~Burnier, A.~Rothkopf,
  \href{https://link.aps.org/doi/10.1103/PhysRevLett.111.182003}{Bayesian
  approach to spectral function reconstruction for euclidean quantum field
  theories}, Phys. Rev. Lett. 111 (2013) 182003.
\newblock \href {https://doi.org/10.1103/PhysRevLett.111.182003}
  {\path{doi:10.1103/PhysRevLett.111.182003}}.
\newline\urlprefix\url{https://link.aps.org/doi/10.1103/PhysRevLett.111.182003}

\bibitem{Sandvik98}
A.~W. Sandvik,
  \href{https://link.aps.org/doi/10.1103/PhysRevB.57.10287}{Stochastic method
  for analytic continuation of quantum monte carlo data}, Phys. Rev. B 57
  (1998) 10287--10290.
\newblock \href {https://doi.org/10.1103/PhysRevB.57.10287}
  {\path{doi:10.1103/PhysRevB.57.10287}}.
\newline\urlprefix\url{https://link.aps.org/doi/10.1103/PhysRevB.57.10287}

\bibitem{Mishchenko00}
A.~S. Mishchenko, N.~V. Prokof'ev, A.~Sakamoto, B.~V. Svistunov,
  \href{https://link.aps.org/doi/10.1103/PhysRevB.62.6317}{Diagrammatic quantum
  monte carlo study of the fr\"ohlich polaron}, Phys. Rev. B 62 (2000)
  6317--6336.
\newblock \href {https://doi.org/10.1103/PhysRevB.62.6317}
  {\path{doi:10.1103/PhysRevB.62.6317}}.
\newline\urlprefix\url{https://link.aps.org/doi/10.1103/PhysRevB.62.6317}

\bibitem{Fuchs10}
S.~Fuchs, T.~Pruschke, M.~Jarrell,
  \href{https://link.aps.org/doi/10.1103/PhysRevE.81.056701}{Analytic
  continuation of quantum monte carlo data by stochastic analytical inference},
  Phys. Rev. E 81 (2010) 056701.
\newblock \href {https://doi.org/10.1103/PhysRevE.81.056701}
  {\path{doi:10.1103/PhysRevE.81.056701}}.
\newline\urlprefix\url{https://link.aps.org/doi/10.1103/PhysRevE.81.056701}

\bibitem{Sandvik16}
A.~W. Sandvik,
  \href{https://link.aps.org/doi/10.1103/PhysRevE.94.063308}{Constrained
  sampling method for analytic continuation}, Phys. Rev. E 94 (2016) 063308.
\newblock \href {https://doi.org/10.1103/PhysRevE.94.063308}
  {\path{doi:10.1103/PhysRevE.94.063308}}.
\newline\urlprefix\url{https://link.aps.org/doi/10.1103/PhysRevE.94.063308}

\bibitem{Krivenko19}
I.~Krivenko, M.~Harland,
  \href{https://www.sciencedirect.com/science/article/pii/S0010465519300402}{Triqs/som:
  Implementation of the stochastic optimization method for analytic
  continuation}, Computer Physics Communications 239 (2019) 166--183.
\newblock \href {https://doi.org/https://doi.org/10.1016/j.cpc.2019.01.021}
  {\path{doi:https://doi.org/10.1016/j.cpc.2019.01.021}}.
\newline\urlprefix\url{https://www.sciencedirect.com/science/article/pii/S0010465519300402}

\bibitem{Shao23}
H.~Shao, A.~W. Sandvik,
  \href{https://www.sciencedirect.com/science/article/pii/S0370157322003921}{Progress
  on stochastic analytic continuation of quantum monte carlo data}, Physics
  Reports 1003 (2023) 1--88, progress on stochastic analytic continuation of
  quantum Monte Carlo data.
\newblock \href {https://doi.org/https://doi.org/10.1016/j.physrep.2022.11.002}
  {\path{doi:https://doi.org/10.1016/j.physrep.2022.11.002}}.
\newline\urlprefix\url{https://www.sciencedirect.com/science/article/pii/S0370157322003921}

\bibitem{Goulko17}
O.~Goulko, A.~S. Mishchenko, L.~Pollet, N.~Prokof'ev, B.~Svistunov,
  \href{https://link.aps.org/doi/10.1103/PhysRevB.95.014102}{Numerical analytic
  continuation: Answers to well-posed questions}, Phys. Rev. B 95 (2017)
  014102.
\newblock \href {https://doi.org/10.1103/PhysRevB.95.014102}
  {\path{doi:10.1103/PhysRevB.95.014102}}.
\newline\urlprefix\url{https://link.aps.org/doi/10.1103/PhysRevB.95.014102}

\bibitem{Otsuki17}
J.~Otsuki, M.~Ohzeki, H.~Shinaoka, K.~Yoshimi,
  \href{https://link.aps.org/doi/10.1103/PhysRevE.95.061302}{Sparse modeling
  approach to analytical continuation of imaginary-time quantum monte carlo
  data}, Phys. Rev. E 95 (2017) 061302.
\newblock \href {https://doi.org/10.1103/PhysRevE.95.061302}
  {\path{doi:10.1103/PhysRevE.95.061302}}.
\newline\urlprefix\url{https://link.aps.org/doi/10.1103/PhysRevE.95.061302}

\bibitem{Yoshimi19}
K.~Yoshimi, J.~Otsuki, Y.~Motoyama, M.~Ohzeki, H.~Shinaoka,
  \href{https://www.sciencedirect.com/science/article/pii/S0010465519302103}{Spm:
  Sparse modeling tool for analytic continuation of imaginary-time green’s
  function}, Computer Physics Communications 244 (2019) 319--323.
\newblock \href {https://doi.org/https://doi.org/10.1016/j.cpc.2019.07.001}
  {\path{doi:https://doi.org/10.1016/j.cpc.2019.07.001}}.
\newline\urlprefix\url{https://www.sciencedirect.com/science/article/pii/S0010465519302103}

\bibitem{Otsuki20}
J.~Otsuki, M.~Ohzeki, H.~Shinaoka, K.~Yoshimi,
  \href{https://doi.org/10.7566/JPSJ.89.012001}{Sparse modeling in quantum
  many-body problems}, Journal of the Physical Society of Japan 89~(1) (2020)
  012001.
\newblock \href {http://arxiv.org/abs/https://doi.org/10.7566/JPSJ.89.012001}
  {\path{arXiv:https://doi.org/10.7566/JPSJ.89.012001}}, \href
  {https://doi.org/10.7566/JPSJ.89.012001} {\path{doi:10.7566/JPSJ.89.012001}}.
\newline\urlprefix\url{https://doi.org/10.7566/JPSJ.89.012001}

\bibitem{Ying22}
L.~Ying,
  \href{https://www.sciencedirect.com/science/article/pii/S0021999122006118}{Analytic
  continuation from limited noisy matsubara data}, Journal of Computational
  Physics 469 (2022) 111549.
\newblock \href {https://doi.org/https://doi.org/10.1016/j.jcp.2022.111549}
  {\path{doi:https://doi.org/10.1016/j.jcp.2022.111549}}.
\newline\urlprefix\url{https://www.sciencedirect.com/science/article/pii/S0021999122006118}

\bibitem{Yoon18}
H.~Yoon, J.-H. Sim, M.~J. Han,
  \href{https://link.aps.org/doi/10.1103/PhysRevB.98.245101}{Analytic
  continuation via domain knowledge free machine learning}, Phys. Rev. B 98
  (2018) 245101.
\newblock \href {https://doi.org/10.1103/PhysRevB.98.245101}
  {\path{doi:10.1103/PhysRevB.98.245101}}.
\newline\urlprefix\url{https://link.aps.org/doi/10.1103/PhysRevB.98.245101}

\bibitem{Fournier20}
R.~Fournier, L.~Wang, O.~V. Yazyev, Q.~Wu,
  \href{https://link.aps.org/doi/10.1103/PhysRevLett.124.056401}{Artificial
  neural network approach to the analytic continuation problem}, Phys. Rev.
  Lett. 124 (2020) 056401.
\newblock \href {https://doi.org/10.1103/PhysRevLett.124.056401}
  {\path{doi:10.1103/PhysRevLett.124.056401}}.
\newline\urlprefix\url{https://link.aps.org/doi/10.1103/PhysRevLett.124.056401}

\bibitem{Huang23}
Z.~Huang, E.~Gull, L.~Lin,
  \href{https://link.aps.org/doi/10.1103/PhysRevB.107.075151}{Robust analytic
  continuation of green's functions via projection, pole estimation, and
  semidefinite relaxation}, Phys. Rev. B 107 (2023) 075151.
\newblock \href {https://doi.org/10.1103/PhysRevB.107.075151}
  {\path{doi:10.1103/PhysRevB.107.075151}}.
\newline\urlprefix\url{https://link.aps.org/doi/10.1103/PhysRevB.107.075151}

\bibitem{Vidberg77}
H.~J. Vidberg, J.~W. Serene, \href{https://doi.org/10.1007/BF00655090}{{Solving
  the Eliashberg equations by means ofN-point Pad{\'{e}} approximants}},
  Journal of Low Temperature Physics 29~(3) (1977) 179--192.
\newblock \href {https://doi.org/10.1007/BF00655090}
  {\path{doi:10.1007/BF00655090}}.
\newline\urlprefix\url{https://doi.org/10.1007/BF00655090}

\bibitem{Beach00}
K.~S.~D. Beach, R.~J. Gooding, F.~Marsiglio,
  \href{https://link.aps.org/doi/10.1103/PhysRevB.61.5147}{Reliable pad\'e
  analytical continuation method based on a high-accuracy symbolic computation
  algorithm}, Phys. Rev. B 61 (2000) 5147--5157.
\newblock \href {https://doi.org/10.1103/PhysRevB.61.5147}
  {\path{doi:10.1103/PhysRevB.61.5147}}.
\newline\urlprefix\url{https://link.aps.org/doi/10.1103/PhysRevB.61.5147}

\bibitem{Kraberger17}
G.~J. Kraberger, R.~Triebl, M.~Zingl, M.~Aichhorn,
  \href{https://link.aps.org/doi/10.1103/PhysRevB.96.155128}{Maximum entropy
  formalism for the analytic continuation of matrix-valued green's functions},
  Phys. Rev. B 96 (2017) 155128.
\newblock \href {https://doi.org/10.1103/PhysRevB.96.155128}
  {\path{doi:10.1103/PhysRevB.96.155128}}.
\newline\urlprefix\url{https://link.aps.org/doi/10.1103/PhysRevB.96.155128}

\bibitem{Rothkopf17}
A.~Rothkopf,
  \href{https://link.aps.org/doi/10.1103/PhysRevD.95.056016}{Bayesian inference
  of nonpositive spectral functions in quantum field theory}, Phys. Rev. D 95
  (2017) 056016.
\newblock \href {https://doi.org/10.1103/PhysRevD.95.056016}
  {\path{doi:10.1103/PhysRevD.95.056016}}.
\newline\urlprefix\url{https://link.aps.org/doi/10.1103/PhysRevD.95.056016}

\bibitem{Gull14}
E.~Gull, A.~J. Millis,
  \href{https://link.aps.org/doi/10.1103/PhysRevB.90.041110}{Pairing glue in
  the two-dimensional hubbard model}, Phys. Rev. B 90 (2014) 041110.
\newblock \href {https://doi.org/10.1103/PhysRevB.90.041110}
  {\path{doi:10.1103/PhysRevB.90.041110}}.
\newline\urlprefix\url{https://link.aps.org/doi/10.1103/PhysRevB.90.041110}

\bibitem{Reymbaut15}
A.~Reymbaut, D.~Bergeron, A.-M.~S. Tremblay,
  \href{https://link.aps.org/doi/10.1103/PhysRevB.92.060509}{Maximum entropy
  analytic continuation for spectral functions with nonpositive spectral
  weight}, Phys. Rev. B 92 (2015) 060509.
\newblock \href {https://doi.org/10.1103/PhysRevB.92.060509}
  {\path{doi:10.1103/PhysRevB.92.060509}}.
\newline\urlprefix\url{https://link.aps.org/doi/10.1103/PhysRevB.92.060509}

\bibitem{Dong22}
X.~Dong, E.~Gull, A.~J. Millis,
  \href{https://doi.org/10.1038/s41567-022-01710-z}{Quantifying the role of
  antiferromagnetic fluctuations in the superconductivity of the doped hubbard
  model}, Nature Physics 18~(11) (2022) 1293--1296.
\newblock \href {https://doi.org/10.1038/s41567-022-01710-z}
  {\path{doi:10.1038/s41567-022-01710-z}}.
\newline\urlprefix\url{https://doi.org/10.1038/s41567-022-01710-z}

\bibitem{Yue23}
C.~Yue, P.~Werner, Maximum entropy analytic continuation of anomalous
  self-energies (2023).
\newblock \href {http://arxiv.org/abs/2303.16888} {\path{arXiv:2303.16888}}.

\bibitem{Fei21}
J.~Fei, C.-N. Yeh, E.~Gull,
  \href{https://link.aps.org/doi/10.1103/PhysRevLett.126.056402}{Nevanlinna
  analytical continuation}, Phys. Rev. Lett. 126 (2021) 056402.
\newblock \href {https://doi.org/10.1103/PhysRevLett.126.056402}
  {\path{doi:10.1103/PhysRevLett.126.056402}}.
\newline\urlprefix\url{https://link.aps.org/doi/10.1103/PhysRevLett.126.056402}

\bibitem{Nevanlinna25}
R.~Nevanlinna, \href{https://doi.org/10.1007/BF02543858}{{Zur Theorie der
  Meromorphen Funktionen}}, Acta Mathematica 46~(1-2) (1925) 1 -- 99.
\newblock \href {https://doi.org/10.1007/BF02543858}
  {\path{doi:10.1007/BF02543858}}.
\newline\urlprefix\url{https://doi.org/10.1007/BF02543858}

\bibitem{Fei21A}
J.~Fei, C.-N. Yeh, D.~Zgid, E.~Gull,
  \href{https://link.aps.org/doi/10.1103/PhysRevB.104.165111}{Analytical
  continuation of matrix-valued functions: Carath\'eodory formalism}, Phys.
  Rev. B 104 (2021) 165111.
\newblock \href {https://doi.org/10.1103/PhysRevB.104.165111}
  {\path{doi:10.1103/PhysRevB.104.165111}}.
\newline\urlprefix\url{https://link.aps.org/doi/10.1103/PhysRevB.104.165111}

\bibitem{Caratheodory1907}
C.~Carath{\'e}odory, {\"Uber den Variabilit\"atsbereich der Koeffizienten von
  Potenzreihen, die gegebene Werte nicht annehmen}, Mathematische Annalen
  64~(1) (1907) 95--115.
\newblock \href {https://doi.org/10.1007/BF01449883}
  {\path{doi:10.1007/BF01449883}}.

\bibitem{Pick17}
G.~Pick, \href{https://link.springer.com/article/10.1007/BF01456817}{Über die
  beschränkungen analytischer funktionen, welche durch vorgegebene
  funktionswerte bewirkt werdenber die beschränkungen analytischer funktionen,
  welche durch vorgegebene funktionswerte bewirkt werden}, Mathematische
  annalen 78 (1917) 270 -- 275.
\newblock \href {https://doi.org/10.1007/BF01457103}
  {\path{doi:10.1007/BF01457103}}.
\newline\urlprefix\url{https://link.springer.com/article/10.1007/BF01456817}

\bibitem{Schur18}
J.~Schur, \"{U}ber potenzreihen, die im innern des einheits-kreises beschränkt
  sind, Journal für die reine und angewandte Mathematik (Crelles Journal) 1918
  (1918) 122 --145.
\newblock \href {https://doi.org/10.1515/crll.1918.148.122}
  {\path{doi:10.1515/crll.1918.148.122}}.

\bibitem{Yeh22B}
C.-N. Yeh, A.~Shee, Q.~Sun, E.~Gull, D.~Zgid,
  \href{https://link.aps.org/doi/10.1103/PhysRevB.106.085121}{Relativistic
  self-consistent $gw$: Exact two-component formalism with one-electron
  approximation for solids}, Phys. Rev. B 106 (2022) 085121.
\newblock \href {https://doi.org/10.1103/PhysRevB.106.085121}
  {\path{doi:10.1103/PhysRevB.106.085121}}.
\newline\urlprefix\url{https://link.aps.org/doi/10.1103/PhysRevB.106.085121}

\bibitem{Bergamaschi23}
T.~Bergamaschi, W.~I. Jay, P.~R. Oare, Hadronic structure, conformal maps, and
  analytic continuation (2023).
\newblock \href {http://arxiv.org/abs/2305.16190} {\path{arXiv:2305.16190}}.

\bibitem{Nogaki23B}
K.~Nogaki, H.~Shinaoka, \href{https://doi.org/10.7566/JPSJ.92.035001}{Bosonic
  nevanlinna analytic continuation}, Journal of the Physical Society of Japan
  92~(3) (2023) 035001.
\newblock \href {http://arxiv.org/abs/https://doi.org/10.7566/JPSJ.92.035001}
  {\path{arXiv:https://doi.org/10.7566/JPSJ.92.035001}}, \href
  {https://doi.org/10.7566/JPSJ.92.035001} {\path{doi:10.7566/JPSJ.92.035001}}.
\newline\urlprefix\url{https://doi.org/10.7566/JPSJ.92.035001}

\bibitem{Nogaki23}
K.~Nogaki, J.~Fei, E.~Gull, H.~Shinaoka, Nevanlinna.jl: A julia implementation
  of nevanlinna analytic continuation (2023).
\newblock \href {http://arxiv.org/abs/2302.10476} {\path{arXiv:2302.10476}}.

\bibitem{43cc8a3d-e842-3b2d-b3bd-f44652881afb}
P.~Delsarte, Y.~Genin, Y.~Kamp, \href{http://www.jstor.org/stable/2100767}{The
  nevanlinna-pick problem for matrix-valued functions}, SIAM Journal on Applied
  Mathematics 36~(1) (1979) 47--61.
\newline\urlprefix\url{http://www.jstor.org/stable/2100767}

\bibitem{doi:10.1137/0602013}
P.~Delsarte, Y.~Genin, Y.~Kamp,
  \href{https://doi.org/10.1137/0602013}{Generalized schur representation of
  matrix-valued functions}, SIAM Journal on Algebraic Discrete Methods 2~(2)
  (1981) 94--107.
\newblock \href {https://doi.org/10.1137/0602013} {\path{doi:10.1137/0602013}}.
\newline\urlprefix\url{https://doi.org/10.1137/0602013}

\bibitem{CHEN1994253}
G.~Chen, Çetin Kaya~Koç,
  \href{https://www.sciencedirect.com/science/article/pii/0024379594902054}{Computing
  matrix-valued nevanlinna-pick interpolation}, Linear Algebra and its
  Applications 203-204 (1994) 253--263.
\newblock \href {https://doi.org/https://doi.org/10.1016/0024-3795(94)90205-4}
  {\path{doi:https://doi.org/10.1016/0024-3795(94)90205-4}}.
\newline\urlprefix\url{https://www.sciencedirect.com/science/article/pii/0024379594902054}

\bibitem{10.1063/1.4826042}
C.~Yazici, H.~K. Sevindir, \href{https://doi.org/10.1063/1.4826042}{{A
  correction for computing matrix-valued Nevanlinna-Pick interpolation
  problem}}, AIP Conference Proceedings 1558~(1) (2013) 2474--2477.
\newblock \href {https://doi.org/10.1063/1.4826042}
  {\path{doi:10.1063/1.4826042}}.
\newline\urlprefix\url{https://doi.org/10.1063/1.4826042}

\bibitem{nevanlinna1929beschrankte}
R.~H. Nevanlinna, R.~Nevanlinna, {\"U}ber beschr{\"a}nkte analytische
  Funktionen, Suomalaisen tiedeakatemian kustantama, 1929.

\bibitem{akhiezer2020classical}
N.~Akhiezer, \href{https://books.google.com/books?id=DhDiEAAAQBAJ}{The
  Classical Moment Problem and Some Related Questions in Analysis}, Classics in
  Applied Mathematics, Society for Industrial and Applied Mathematics, 2020.
\newline\urlprefix\url{https://books.google.com/books?id=DhDiEAAAQBAJ}

\bibitem{10.1155/S1687120003212028}
V.~M. Adamyan, J.~Alcober, I.~M. Tkachenko,
  \href{https://doi.org/10.1155/S1687120003212028}{{Reconstruction of
  distributions by their moments and local constraints}}, Applied Mathematics
  Research eXpress 2003~(2) (2003) 33--70.
\newblock \href {https://doi.org/10.1155/S1687120003212028}
  {\path{doi:10.1155/S1687120003212028}}.
\newline\urlprefix\url{https://doi.org/10.1155/S1687120003212028}

\bibitem{TriqsNevDocu}
S.~Iskakov, A.~Hampel, N.~Wentzell, E.~Gull,
  https://triqs.github.io/Nevanlinna/latest/ (2023).
\newblock \href{https://triqs.github.io/Nevanlinna/latest/}{[link]}.
\newline\urlprefix\url{https://triqs.github.io/Nevanlinna/latest/}

\bibitem{SCIPY}
E.~Jones, T.~Oliphant, P.~Peterson, et~al.,
  \href{http://www.scipy.org/}{{SciPy}: Open source scientific tools for
  {Python}} (2001--).
\newline\urlprefix\url{http://www.scipy.org/}

\bibitem{Nocedal2006}
\href{https://doi.org/10.1007/978-0-387-40065-5_5}{Conjugate Gradient Methods},
  Springer New York, New York, NY, 2006, pp. 101--134.
\newblock \href {https://doi.org/10.1007/978-0-387-40065-5_5}
  {\path{doi:10.1007/978-0-387-40065-5_5}}.
\newline\urlprefix\url{https://doi.org/10.1007/978-0-387-40065-5_5}

\end{thebibliography}

\end{document}